\documentclass[useAMS]{mn2e}

\usepackage{times}
\usepackage{graphicx}
\usepackage{amsmath}
\usepackage{amssymb}
\usepackage{color}
\newcommand{\beq}{\begin{equation}}
\newcommand{\bea}{\begin{eqnarray}}
\newcommand{\eeq}{\end{equation}}
\newcommand{\eea}{\end{eqnarray}}

\title[Current filamentation upstream of relativistic
  shocks]{Current-driven filamentation upstream of magnetized
  relativistic collisionless shocks}

\author[M. Lemoine, G. Pelletier, L. Gremillet \& I. Plotnikov]{
  Martin Lemoine$^1$\thanks{e-mail:{\tt lemoine@iap.fr}},
  Guy Pelletier$^2$\thanks{e-mail:{\tt
      guy.pelletier@obs.ujf-grenoble.fr}},
  Laurent Gremillet$^3$\thanks{e-mail:{\tt laurent.gremillet@cea.fr}},
  Illya Plotnikov$^2$\thanks{e-mail:{\tt
      illya.plotnikov@obs.ujf-grenoble.fr}}\\
  $^1$ Institut d'Astrophysique de Paris, CNRS, UPMC,
  98 bis boulevard Arago, F-75014 Paris, France\\
  $^2$ UJF-Grenoble 1 / CNRS-INSU, Institut de Plan\'etologie et
  d'Astrophysique de Grenoble (IPAG) UMR 5274,
  F-38041 Grenoble, France\\
  $^3$ CEA, DAM, DIF, F-91297 Arpajon, France 
}

\begin{document}

\date{}

\pubyear{2013}

\maketitle

\label{firstpage}

\begin{abstract}
  The physics of instabilities in the precursor of relativistic
  collisionless shocks is of broad importance in high energy
  astrophysics, because these instabilities build up the shock,
  control the particle acceleration process and generate the magnetic
  fields in which the accelerated particles radiate. Two crucial
  parameters control the micro-physics of these shocks: the
  magnetization of the ambient medium and the Lorentz factor of the
  shock front; as of today, much of this parameter space remains to be
  explored. In the present paper, we report on a new instability
  upstream of electron-positron relativistic shocks and we argue that
  this instability shapes the micro-physics at moderate magnetization
  levels and/or large Lorentz factors. This instability is seeded by
  the electric current carried by the accelerated particles in the
  shock precursor as they gyrate around the background magnetic
  field. The compensation current induced in the background plasma
  leads to an unstable configuration, with the appearance of charge
  neutral filaments carrying a current of the same polarity, oriented
  along the perpendicular current. This ``current-driven
  filamentation'' instability grows faster than any other instability
  studied so far upstream of relativistic shocks, with a growth rate
  comparable to the plasma frequency. Furthermore, the compensation of
  the current is associated with a slow-down of the ambient plasma as
  it penetrates the shock precursor (as viewed in the shock rest
  frame). This slow-down of the plasma implies that the ``current
  driven filamentation'' instability can grow for any value of the
  shock Lorentz factor, provided the magnetization
  $\sigma\,\lesssim\,10^{-2}$. We argue that this instability explains
  the results of recent particle-in-cell simulations in the mildly
  magnetized regime.
\end{abstract}

\begin{keywords} 
Acceleration of particles -- Shock waves 
\end{keywords}

\section{Introduction}\label{sec:introd}

The physics of particle acceleration at relativistic collisionless
shock waves plays a key role in the description of a number of powerful
astrophysical objects, e.g. blazars, pulsar wind nebulae, gamma-ray
bursts etc. One of the lessons learned in the past decade in this field of
research, is the importance of the non-linear relationship that ties
the acceleration process and the generation of micro-turbulence in the
shock vicinity. It was anticipated early on that the self-generation
of micro-turbulence on length scales much smaller than the gyroradius
of the accelerated particles is a necessary condition for the proper
development of the relativistic Fermi process (Lemoine et al. 2006),
in agreement with test particle Monte Carlo simulations (Niemiec et
al. 2006). This small-scale nature of the turbulence comes with a
number of important consequences, most notably the limited maximal
energy of particles accelerated at ultra-relativistic shock waves,
e.g. Kirk \& Reville (2010), Bykov et al. (2012), Plotnikov et
al. (2013a). 

The particle-in-cell (PIC) numerical simulations of Spitkovsky
(2008a,b) have confirmed the validity of these arguments and offered a
more exhaustive picture of the acceleration process in the
ultra-relativistic unmagnetized limit. These simulations have shown
that the accelerated (supra-thermal) particle population excites
filamentation instabilities upstream of unmagnetized shock waves
(meaning, shock waves propagating in an unmagnetized medium), see also
Nishikawa et al. (2009); these instabilities build up a magnetic
barrier on plasma scales $c/\omega_{\rm p}$ and at the same time serve
as scattering centers for the acceleration process. As the magnetic
field energy density grows to an equipartition fraction $\epsilon_B
\,\sim\,10^{-1}$ ($\epsilon_B$ denotes the fraction of incoming
kinetic energy flux in the shock front rest frame stored in magnetic
energy), incoming particles can be isotropized on a coherence length
scale of the order of $\sim 10c/\omega_{\rm p}$, thereby initiating
the shock transition. The gyroradius of accelerated particles remains
larger than this length scale and the Fermi acceleration process
develops as anticipated. These simulations have been confirmed, and
followed by further PIC simulations with different conditions, in
particular regarding the degree of magnetization of the upstream
(background) plasma, the obliquity of the magnetic field and the
nature (pairs vs electron-proton) of the incoming flow (e.g. Keshet et
al. 2009, Martins et al. 2009, Sironi \& Spitkovsky 2009, 2011,
Haugb\o lle 2011, Sironi et al. 2013).

The physics of the electromagnetic instabilities that lead to the
formation of a ultra-relativistic collisionless shock and to the
self-sustainance of the shock have naturally received a lot of
attention: e.g. Hoshino \& Arons (1991), Hoshino et al. (1992) and
Gallant et al. (1992) for magnetized shock waves; for weakly
magnetized shock waves, see e.g.  Medvedev \& Loeb (1999), Wiersma \&
Achterberg (2004), Lyubarsky \& Eichler (2006), Milosavljevi\'c \&
Nakar (2006), Achterberg \& Wiersma (2007), Achterberg et al. (2007),
Pelletier et al. (2009), Lemoine \& Pelletier (2010, 2011), Bret et
al. (2010), Rabinak et al. (2011) and Shaisultanov et al. (2012). To
summarize in a few lines the current understanding, the
Weibel/filamentation instability appears to play a leading role in the
generation of the small-scale magnetic field in the weakly magnetized
shock limit, although electrostatic oblique modes and Buneman modes
retain their importance in pre-heating the electrons away from the
shock front; see the discussion in Lemoine \& Pelletier (2011). At
strongly magnetized shock waves, the synchrotron maser instability is
recognized as the leading agent of dissipation, e.g.  Hoshino \& Arons
(1991), Hoshino et al. (1992) and Gallant et al. (1992).

However, at intermediate magnetizations and/or very large Lorentz
factors, the physics remains poorly known. Indeed, the filamentation
instability and other two stream modes cannot be excited in these
regions of parameter space, because the timescale on which the
incoming particles cross the precursor becomes shorter than the
timescale on which such instabilities can be excited (Lemoine \&
Pelletier 2010, 2011). Therefore, how the shock is structured in such
conditions remains an open question.

We report here on a new current-driven instability which is likely to
emerge as the dominant instability in this range of magnetization and
at very large Lorentz factors. The electric current is carried by the
suprathermal particles (or shock reflected particles) and results from
their gyration in the background magnetic field: assuming that the
magnetic field is oriented along the $\boldsymbol{z}$ axis, while the
incoming plasma flows along $\boldsymbol{-x}$ in the shock rest frame,
the current is generated along $\boldsymbol{-y}$, since the Lorentz
force deflects positive and negative suprathermal particles in
opposite directions. As the ambient plasma penetrates the precursor,
it develops a compensating current along $\boldsymbol{+y}$. This
configuration is found to be unstable, because a current fluctuation
can couple to a density fluctuation and excite a combination of
extraordinary modes and compressive modes of the ambient plasma. This
will be made explicit further on.

As viewed from the rest frame of the ambient plasma, this
perpendicular electric current is extraordinarily large. If one writes
$\xi_{\rm cr}$ the fraction of incoming kinetic energy flux carried by
the suprathermal particles -- see Eq.~(\ref{eq:xi}) below -- with
$\xi_{\rm cr}\sim 0.1$ indicated by PIC simulations, $\gamma_{\rm
  sh}\,\gg\,1$ the Lorentz factor of the shock wave in the ambient
plasma frame and $n_{\rm u}$ the proper density of the ambient plasma,
the induced current reads $j_{y,\rm cr} \,\sim\, \gamma_{\rm
  sh}\xi_{\rm cr} n_{\rm u} e c$. For $\gamma_{\rm sh}\xi_{\rm
  cr}\,\gtrsim\,1$, as expected in ultra-relativistic shocks, this
current \emph{cannot} be compensated by the ambient plasma at rest. As
we will demonstrate, the latter is actually accelerated to
relativistic velocities relatively to its initial rest frame and it is
squashed to an apparent density $\sim \gamma_{\rm sh}\xi_{\rm cr}
n_{\rm u}$ in the frame in which there is no bulk motion along
$\boldsymbol{x}$ (denoted ${\cal R}$ in the following); then, particle
motion at relativistic velocities along $\boldsymbol{y}$ leads to
current compensation.

In this work, we focus on an electron-positron shock; in electron-ion
shocks, a similar current develops but excites other modes, in
particular Whistler waves. This case will be discussed in a
forthcoming paper. In Section~\ref{sec:linear}, we discuss the physics
of the instability at the linear level, using a relativistic two-fluid
model for the incoming background plasma exposed to a rigid external
current set by the suprathermal particles. In Section~\ref{sec:disc},
we discuss the relevance of this instability in relativistic
collisionless shocks and compare it to results of recent PIC
simulations. We discuss the structure of the precursor in
Appendix~\ref{sec:app} and provide conclusions in Sec.~\ref{sec:conc}.

\section{Current-driven filamentation instability}\label{sec:linear}
We describe the shock precursor as follows, in the shock front
frame. The incoming plasma flows with 4-velocity $u_x<0$, carrying
magnetic field $\boldsymbol{B}=B_z\,\boldsymbol{z}$ and convective
electric field $\boldsymbol{E}=\gamma_{\rm sh}\beta_{\rm sh}B_{\rm
  u}\,\boldsymbol{y}$, with $\beta_{\rm sh}<0$ the velocity of the
incoming background plasma in the shock rest frame in units of $c$,
i.e. $\gamma_{\rm sh}\,\equiv\,\left(1-\beta_{\rm
  sh}^2\right)^{-1/2}$. In principle, $B_z$ depends on $x$, while
$B_{\rm u}$ corresponds to the upstream magnetic field measured in the
upstream rest frame well beyond the precursor. The precursor also
contains a population of relativistic suprathermal particles, which
rotate around $\boldsymbol{B}$ and thereby induce a current along
$\boldsymbol{y}$, $\boldsymbol{j_{\rm cr}}\sim -\gamma_{\rm
  sh}\xi_{\rm cr} n_{\rm u}e c\,\boldsymbol{y}$.  The quantity
$\xi_{\rm cr}$ characterizes the fraction of the incoming particle
energy carried by the suprathermal particles:
\begin{equation}
\xi_{\rm cr}\,\equiv\, \frac{e_{\rm cr}}{\gamma_{\rm sh}^2 n_{\rm u}m
  c^2}\ ,\label{eq:xi}
\end{equation}
with $e_{\rm cr} \,=\, n_{\rm cr}\gamma_{\rm sh}m c^2$ in the shock
frame, assuming that the supra-thermal particles carry a density
$n_{\rm cr}$ and typical Lorentz factor $\gamma_{\rm sh}$; from
Eq.~(\ref{eq:xi}), one derives $n_{\rm cr}\,=\,\gamma_{\rm sh}\xi_{\rm
  cr}n_{\rm u}$, whence the expression for the current density
$\boldsymbol{j_{\rm cr}}$.

\begin{figure}
\includegraphics[bb=40 90 570 520, width = 0.48\textwidth]{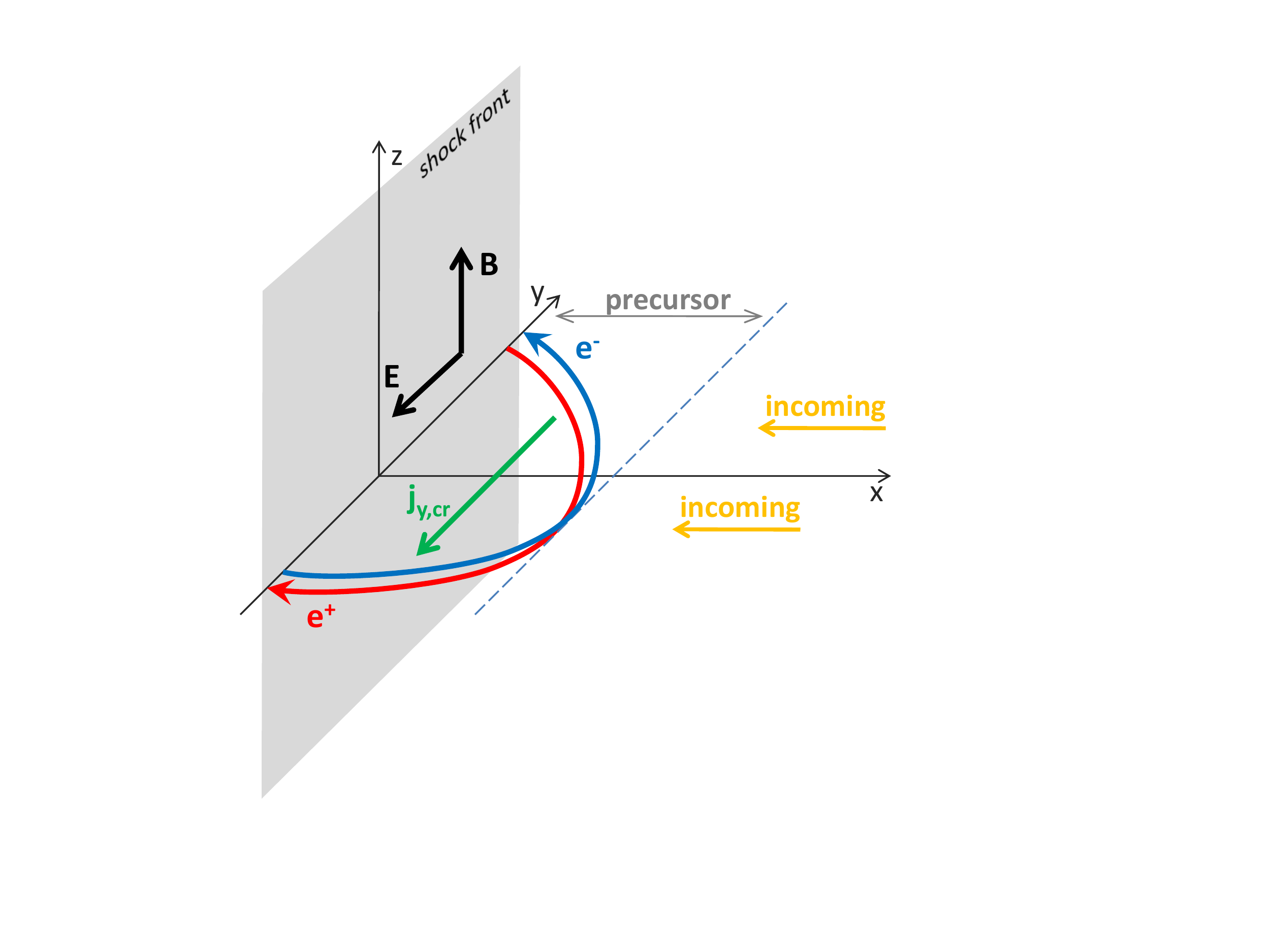}
\caption{Sketch of the precursor of a relativistic magnetized pair
  shock, as viewed in the shock front rest frame.  The
  reflected/shock-accelerated suprathermal particles (in red and blue)
  gyrate in the background magnetic field and accelerate parallel to
  the convective electric field, thereby generating a net
  perpendicular current $j_{y,\rm cr}$. The incoming plasma must
  compensate this current as it penetrates the
  precursor.\label{fig:skp} }
\end{figure}

The spatial profile of this current and the overall structure of the
precursor are described in detail in App.~\ref{sec:app};
Fig.~\ref{fig:skp} offers a sketch of the precursor. The typical size
of the precursor is $c/\omega_{\rm c}$, with $\omega_{\rm c}=eB_{\rm
  u}/(mc)$ the upstream cyclotron frequency; this size also
corresponds to the typical gyration radius $r_{\rm L}$ of the
suprathermal particles in the shock front rest frame, whose typical
Lorentz factor $\sim\gamma_{\rm sh}$.

As the incoming particles cross the precursor, they are deflected
along $\boldsymbol{y}$ in order to compensate the cosmic ray
perpendicular current. Positrons drift towards $\boldsymbol{+y}$ while
electrons drift towards $\boldsymbol{-y}$. The absolute value of the
4-velocity $y-$component for both fluids is equal, $\vert u_y\vert \sim
\gamma_{\rm sh}\xi_{\rm cr}$ (in units of c), hence $\vert u_y\vert\,\gtrsim\,1$
is expected for relativistic shocks, possibly $\vert u_y\vert\,\gg\,1$.

The deflection of the incoming flow along $\boldsymbol{y}$ implies a
substantial deceleration of the flow along $\boldsymbol{x}$, which has
drastic consequences regarding the development of the instability. The
profile of the velocity of the flow is discussed in detail in
App.~\ref{sec:app}, but one can apprehend this slow-down as follows:
the total Lorentz factor of the flow remains large, in particular the
total 3-velocity $\vert\boldsymbol{\beta}\vert\sim 1$, up to
corrections of order $\gamma_{\rm sh}^{-2}$; however, a transverse
velocity develops with magnitude $\vert\beta_y\vert \,\simeq\,\xi_{\rm
  cr}$; the combination of these two facts implies that $\beta_{\rm
  x}$ deviates from unity by quantities of order $\gamma_{\rm
  sh}^{-2}$ or $\xi_{\rm cr}^2$, whichever is larger. In other words,
assuming that $\gamma_{\rm sh}\xi_{\rm cr}\,\gg\,1$, as expected in
ultra-relativistic shocks, leads to $\vert\beta_{\rm
  x}\vert\,\simeq\,1-\xi_{\rm cr}^2/2$. If $\gamma_{\rm sh}\xi_{\rm
  cr}\,\ll\,1$, $\beta_{\rm x}$ remains unchanged compared to the
asymptotic value outside the precursor.

This is a quite remarkable feature: the compensation of the current
slows down the incoming plasma down to the (longitudinal) velocity
$\beta_{\rm x}$; thus, the ${\cal R}$ frame which corresponds to the
instantaneous rest frame of the plasma, in which there is no bulk
motion along $\boldsymbol{x}$, moves with velocity $\beta_{{\cal
    R}\vert\rm sh} = \beta_{\rm x}$ relative to the shock front rest
frame. At large values of the current, $\gamma_{\rm sh}\xi_{\rm
  cr}\,\gg\,1$, the relative Lorentz factor between the ${\cal R}$
frame and the shock front rest frame becomes of the order of
$1/\xi_{\rm cr}$, independent of the far upstream Lorentz factor. In
this sense, the shock precursor plays the role of a buffer, with
important consequences for the physics of the shock, discussed in
Sec.~\ref{sec:disc}.

The Lorentz factor that corresponds to the relative velocity between
this new rest frame ${\cal R}$ and the \emph{far upstream} rest frame
is easily calculated and well approximated by:
\begin{equation}
\gamma_{{\cal R}\vert\rm u}\,\simeq\, {\rm max}\left(1, \gamma_{\rm
    sh}\xi_{\rm cr}/2\right)\ . \label{eq:gru}
\end{equation}

In the following, we analyze the evolution of the instability in the
linear regime by adopting a relativistic two-fluid description of the
incoming plasma, where two-fluid refers to the electron and positron
components of the background plasma. This means, in particular, that
we neglect the response of the cosmic rays and we treat as external
the current that these suprathermal particles carry. The latter
assumption is discussed in Sec.~\ref{sec:disc}. In this section, we
assume that current compensation is achieved to high accuracy in the
shock precursor, as motivated by our discussion in Sec.~\ref{sec:MHD};
see also the discussion in Sec.~5.3.1 of Lemoine \& Pelletier (2011).
This two-fluid description allows us to probe the physics of the
instability up to the inertial scale of the incoming plasma, where the
growth rate is found to peak.

We write and solve the system in the instantaneous rest frame ${\cal
  R}$ of the plasma, in which there is no bulk motion along
$\boldsymbol{x}$. In such a rest frame, the instability is expected to
be absolute (vs convective), provided the growth rate exceeds the
inverse crossing time of the precursor. In the ${\cal R}$ frame,
$u_{x\vert{\cal R}}\,=\,0$ (henceforth, all quantities concern the
incoming plasma), but the (unperturbed) background electric and
magnetic fields read
\begin{equation}
B_{z\vert{\cal R}}\,=\,\gamma_{{\cal R}\vert\rm u} B_{\rm u},
\quad
E_{y\vert{\cal R}}\,=\,-\gamma_{{\cal  R}\vert\rm u}\beta_{{\cal
    R}\vert\rm u} B_{\rm u}\ .
\end{equation}

\subsection{Linear analysis}\label{sec:sysa}
For simplicity, we assume the plasma and the velocity profile to be
uniform throughout the precursor. It is possible to incorporate the
terms associated to the variation of the profile by writing the system
first in the shock front frame, then boosting it to the instantaneous
rest frame of the incoming plasma. The new terms that appear contain
spatial derivatives (along $x$) of the various unperturbed
quantities. The typical magnitude of these inhomegeneous terms
relative to the other terms is of order $\omega_{\rm c}/\omega$ in
Fourier variables; therefore, the above assumption will be justified
provided $\vert k_x\vert\,\gg\,\omega_{\rm c}/c$. As we show in the
following, the growth rate peaks at values close to $\omega_{\rm p}$
on short wavelengths, i.e. $k\,\simeq\,\omega_{\rm p}/c$; this
therefore justifies the above approximation of a uniform precursor.

Our linear analysis is based on a relativistic two-fluid model of the
background plasma subject to the external current imposed by the
gyrating supra-thermal particles. We thus perturb all variables of the
incoming flow and the electromagnetic structure. The unperturbed
equations are:
\begin{eqnarray}
\partial_\mu\left(n_{\pm} u_{\pm}^\mu\right)&\,=\,&0\nonumber\\
\partial_\mu T^{\mu\nu}_{\pm}&\,=\,& \pm e n_{\pm}u_{\pm}^\mu\,F^{\nu}_{\,\,\mu}\ .
\end{eqnarray}
The indices $\pm$ refer to the positron/electron species of the
background plasma, $u^\mu_{\pm}$ to the $4-$velocity and
$T^{\mu\nu}_{\pm}$ to the corresponding energy-momentum tensors. The
perturbed system then reads:
\begin{eqnarray}
  u^\mu_{\pm}\partial_\mu \left(\frac{\delta n_{\pm}}{n}\right) +
  \partial_\mu\delta u^\mu_{\pm}&\,=\,&0\nonumber\\
  u_{\pm}^\mu\partial_\mu \delta u^\nu_{\pm} + \beta_{\rm
    s}^2\partial^\nu\left(\frac{\delta n_{\pm}}{n}\right)
  &\,=\,&\pm
  \frac{e}{m}\delta u^\sigma_{\pm} F^\nu_{\,\,\sigma}  \pm \frac{e}{m}
  u^\sigma_{\pm} \delta F^\nu_{\,\,\sigma} \ ,\nonumber\\\label{eq:sys0}
  &&
\end{eqnarray}
together with the Maxwell equations. We have implicitly assumed a cold
background plasma limit, although we incorporate temperature effects
through the sound velocity $\beta_{\rm s}$.

We recombine the two fluid variables $\delta n_{\pm}$ and $\delta
u_{\pm}^\mu$ into:\footnote{We use a metric with signature
  $(-,+,+,+)$.}
\begin{eqnarray}
  \delta n &\,\equiv\,& \frac{\delta n_+ + \delta n_-}{2},\quad
  \delta \rho \,\equiv\, \frac{\delta n_+ - \delta n_-}{2} \\
  \delta u^\mu &\,\equiv\,& \frac{\delta u_+^\mu + \delta u_-^\mu}{2},\quad
  \Delta u^\mu \,\equiv\, \frac{\delta u_+^\mu - \delta u_-^\mu}{2} \ .\label{eq:dn}
\end{eqnarray}

Of course, to zeroth order, $n_-=n_+\equiv n$, $u_-^0=u_+^0\equiv
u^0$, $u_{+,y}=-u_{-,y}\equiv u_y$. Furthermore,
$\left(u_\pm^\mu+\delta u_{\pm}^\mu\right)\left(u_{\pm\mu}+\delta
  u_{\pm\mu}\right)=-1$ implies
\begin{equation}
  \delta u^0\,=\, \beta_y u^0\Delta
  u_y\ ,\quad
  \Delta u^0\,=\, \beta_y u^0\delta
  u_y\ ,
\end{equation}
with $\beta_y\,\equiv\,u_y/u^0$.  In the ${\cal R}$ frame, in which we
are working here, $u^0\,=\,(1+u_y^2)^{1/2}$; therefore
$u_y\,\sim\,\gamma_{\rm sh}\xi_{\rm cr}\,\gg\,1$ at large shock
Lorentz factors implies $\vert\beta_y\vert\,\sim\,1$. In the limit
$\gamma_{\rm sh}\xi_{\rm cr}\,\gg\,1$ (but $\xi_{\rm cr}\,\ll\,1$),
the parameters $\gamma_{{\cal R}\vert\rm u}/u^0\,\simeq\,1/2$ and
$\beta_{{\cal R}\vert\rm u}\,\simeq\,1$.

The perturbed current $\delta j^\mu=\delta j^\mu_+ + \delta j^\mu_-$
reads
\begin{eqnarray}
  \delta j^0 &\,=\,& 2n ec \,\left(\Delta u^0 +  u^0\delta \rho/n\right), \\
  \delta j_x &\,=\,& 2n ec \,\Delta u_x , \\
  \delta j_y &\,=\,& 2n ec \,\left(\Delta u_y + \beta_y u^0\delta n/n\right), \\
  \delta j_z &\,=\,& 2n ec \,\Delta u_z\ .
\end{eqnarray}

We define the plasma frequency following:
$\omega_p^2\,=\,\omega_{p+}^2 + \omega_{p-}^2\,=\,8\pi n e^2/m_e$,
and the magnetization parameter:
\begin{equation}
\sigma\,=\, \frac{B_{\rm u}^2}{8\pi n m_e c^2}\,=\,\frac{\omega_{\rm
    c}^2}{\omega_{\rm p}^2}\ .\label{eq:sigma}
\end{equation}

The full dispersion relation is calculated from the linear system
discussed in App.~\ref{sec:sys}, by going through Fourier variables,
then taking the determinant of the matrix using the Mathematica
package. This dispersion relation is too lengthy to be reported
here. 

\begin{figure}
\includegraphics[width = 0.48\textwidth]{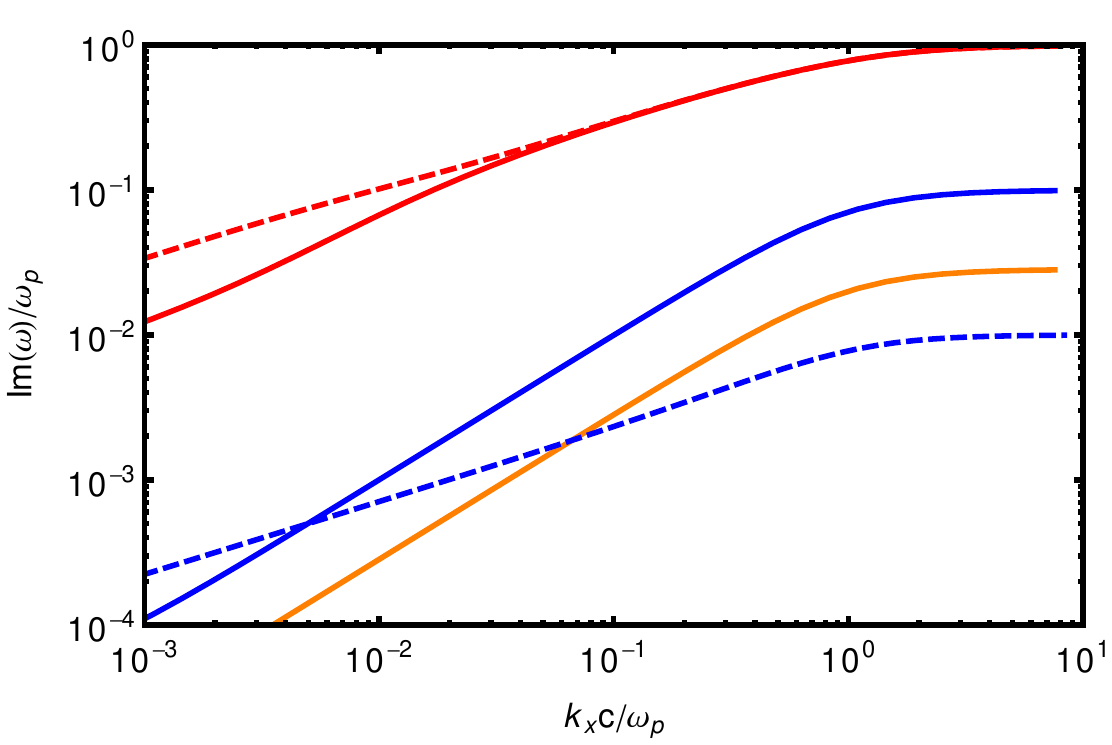}
\caption{Growth rate $\Im\omega/\omega_{\rm p}$ vs $k_x c/\omega_{\rm
    p}$ for $k_y=k_z=0$. In solid lines, $\sigma=10^{-4}$, $\beta_{\rm
    s}=0$ and from top to bottom, $\beta_y=0.99$ (red, corresponding
  to $u_y=\gamma_{\rm sh}\xi_{\rm cr}\,\simeq\,7.0$), $\beta_y=0.1$ (blue,
  corresponding to $\gamma_{\rm sh}\xi_{\rm cr}=0.1$), and
  $\beta_y=0.03$ (orange, $\gamma_{\rm sh}\xi_{\rm cr}=0.03$). In
  dashed lines, same as above for $\sigma=10^{-2}$ (the growth is
  strongly suppressed for $\beta_y=0.03$ in this case).
\label{fig:inst1d} }
\end{figure}

However, it can be given in the following form in the 1D approximation
$k_y=k_z=0$, cold plasma limit $\beta_{\rm s}=0$:
\begin{equation}
\begin{split}
  \omega^5 + i \beta_{{\cal R}\vert\rm u}\beta_y \kappa \omega_{\rm
    c}\omega^4 \nonumber\\- \left[k_x^2c^2 + \kappa^2 \omega_{\rm c}^2
    +(1-\beta_y^2)\omega_{\rm p}^2\right]\omega^3 - i \beta_{{\cal
      R}\vert\rm u}\beta_y \kappa k_x^2c^2\omega_{\rm
    c}\omega^2 \nonumber\\
  + \left(\kappa^2k_x^2c^2\omega_{\rm c}^2 - \beta_y^2 k_x^2c^2
    \omega_{\rm p}^2 -i \kappa \beta_y^3 k_x c\omega_{\rm
      c}\omega_{\rm p}^2\right)\omega \nonumber\\ - i \beta_{{\cal
      R}\vert\rm u}\beta_y^3 \kappa k_x^2c^2\omega_{\rm c}\omega_{\rm
    p}^2\,=\,0
  \nonumber\\
\end{split}
\end{equation}
We recall here the definition $\kappa\,\equiv\, \gamma_{{\cal
    R}\vert\rm u}/u^0$, see App.~\ref{sec:sys}. The growth rate is
represented as a function of $k_x$ for various values of the
parameters $\beta_y$ and $\sigma$ in Fig.~\ref{fig:inst1d}. The global
trend that emerges is a maximal growth rate
\begin{equation}
\Im\omega \,\sim\, \beta_y \omega_{\rm p} \quad
\left(k_xc\,\sim\,\omega_{\rm p},
  \,\,\beta_y\,\gg\,\sqrt{\sigma},\,\,\beta_y\,\gg\,\beta_{\rm
    s}\right)\ .
\end{equation}
The growth rate collapses as soon as one of the conditions indicated
in the brackets is no longer satisfied. The last condition
$\beta_y\,\gg\,\beta_{\rm s}$ is typical of current-driven
instabilities: as the temperature rises and the thermal velocity
exceeds the drift velocity, the instability disappears. However, we do
not expect this situation in ultra-relativistic pair shocks with
$\gamma_{\rm sh}\xi_{\rm cr}\,\gg\,1$, since $\beta_y\,\sim\,1$ in
that limit, while the heating of the incoming flow inside the
precursor remains limited to sub-relativistic velocities, see
e.g. Lemoine \& Pelletier (2011) for a discussion and Spitkovsky
(2008a) for PIC simulations.

In the 2D $k_y=0$, cold plasma ($\beta_{\rm s}=0$), and small current
limit ($\xi_{\rm cr}\gamma_{\rm sh}\,\ll\,1$, in which case
$\beta_{{\cal R}\vert\rm u}\,\sim\,0$ and $u^0\,\sim\,1$), the
dispersion relation also reduces to the compact form:
\begin{eqnarray}
&&\omega^6 -\omega^4\left(\omega_{\rm p}^2+\omega_{\rm
    c}^2+k_x^2c^2+k_z^2c^2-\beta_y^2\omega_{\rm p}^2\right)\nonumber\\
&&+\,\omega^2\left[(k_x^2+k_z^2)c^2\omega_{\rm c}^2 - \beta_y^2 (k_x^2+k_z^2)c^2 \omega_{\rm
    p}^2  - i \beta_y^3 k_xc \omega_{\rm p}^2\omega_{\rm c}\right]\nonumber\\
&&+\, \beta_y^2 k_z^2c^2\omega_{\rm c}^2\omega_{\rm p}^2\,=\,0 \ .\label{eq:disp1} 
\end{eqnarray}
In this limit, the instability can be shown to result from a coupling
between the high frequency branch of the extraordinary mode with the
acoustic mode, as discussed in the following Sec.~\ref{sec:ana}.

\begin{figure}
\includegraphics[width =0.48\textwidth]{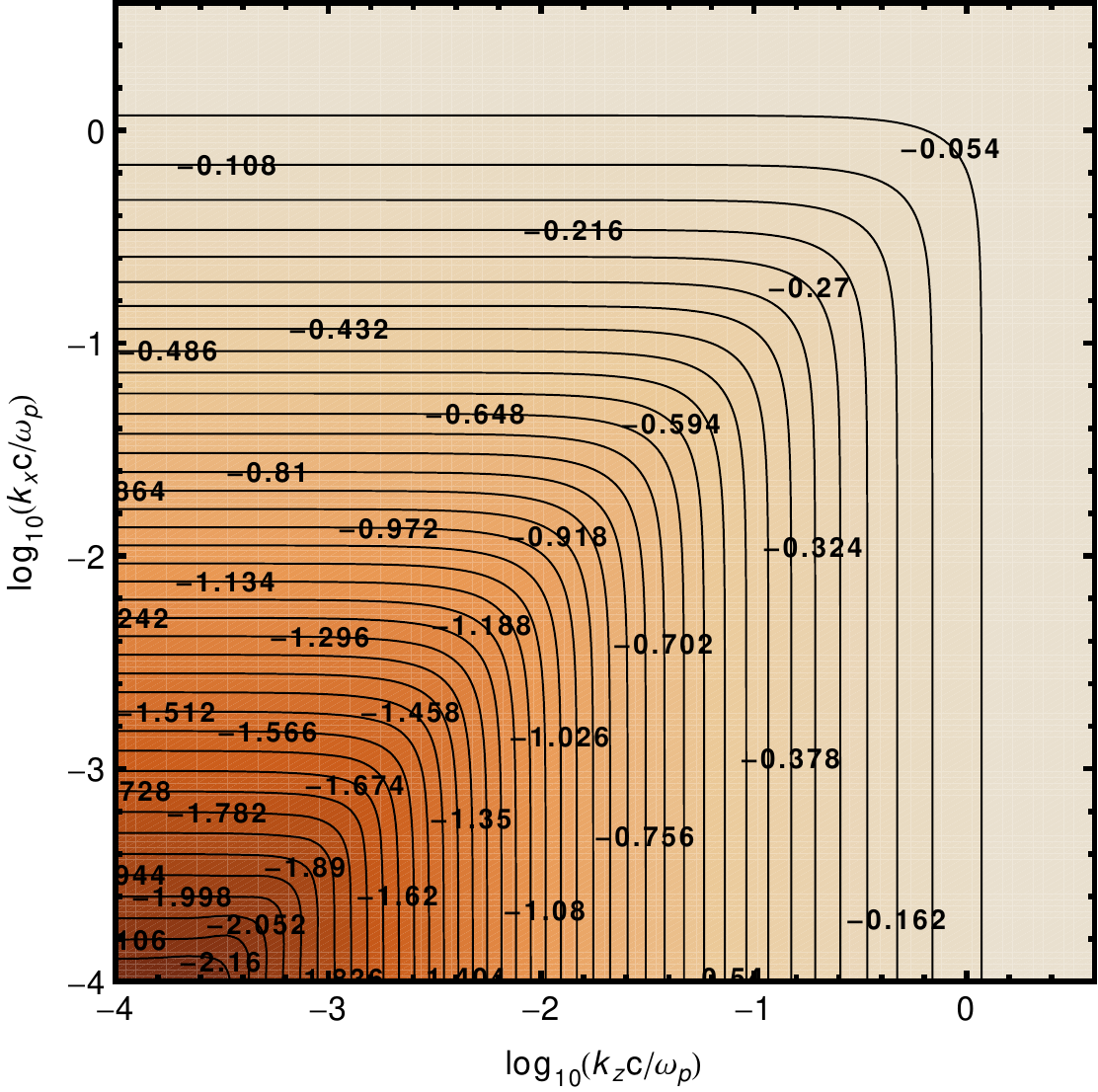}
\caption{Contour plot of ${\rm log}_{10}\left(\Im\omega/\omega_{\rm
      p}\right)$ assuming $k_y\,=\,0$, for $\gamma_{\rm sh}u_y\,=\,7$
  (i.e. $\beta_y\,=\,0.99$),  $\sigma=10^{-3}$, $\beta_{\rm s}\,=\,0$.
\label{fig:inst2dxz} }
\end{figure}

\begin{figure}
\includegraphics[width = 0.48\textwidth]{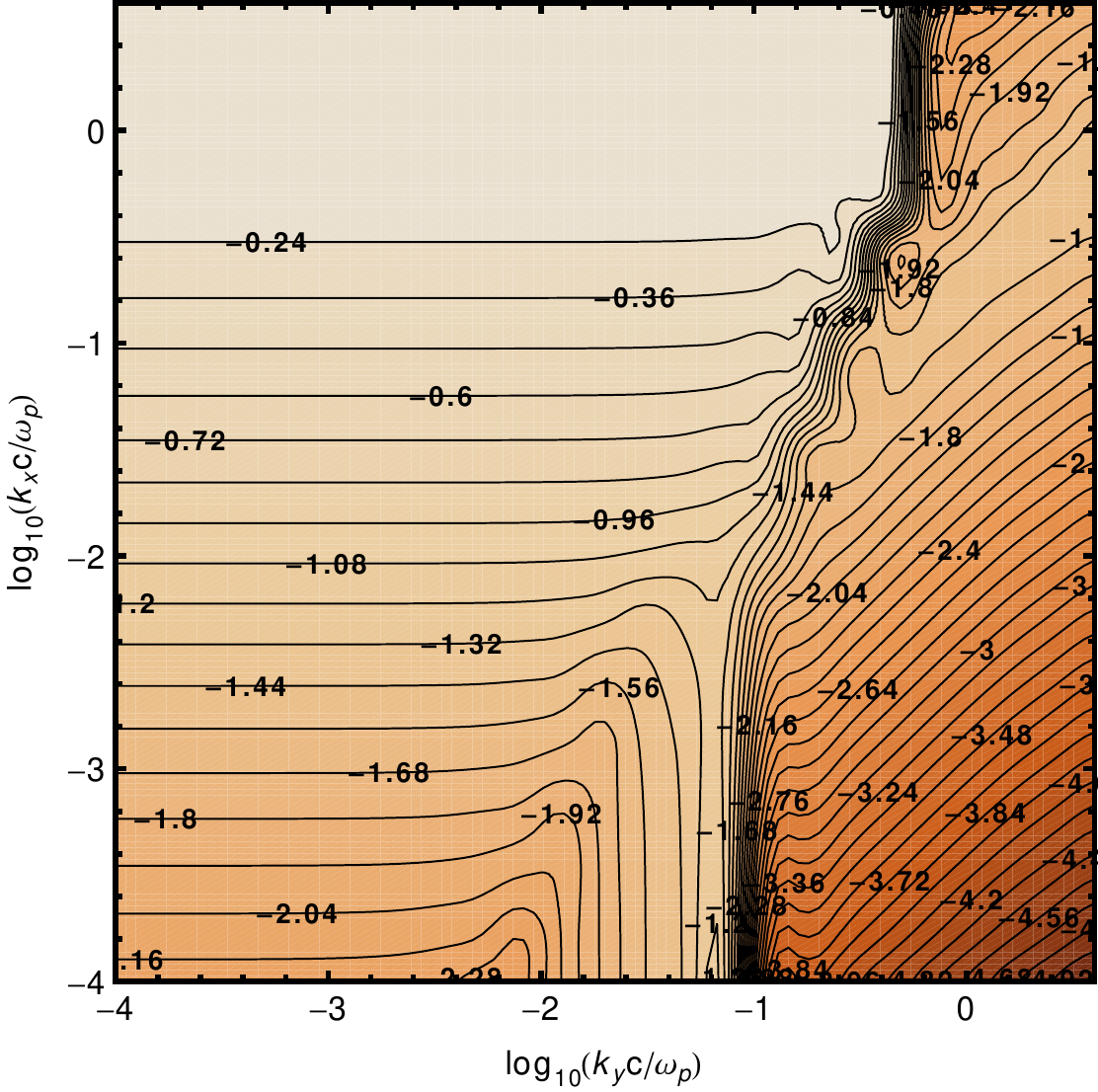}
\caption{Same as Fig.~\ref{fig:inst2dxz}, in the plane $(k_x,k_y)$,
  for $k_z=0$.\label{fig:inst2dxy} }
\end{figure}

\begin{figure}
\includegraphics[width = 0.48\textwidth]{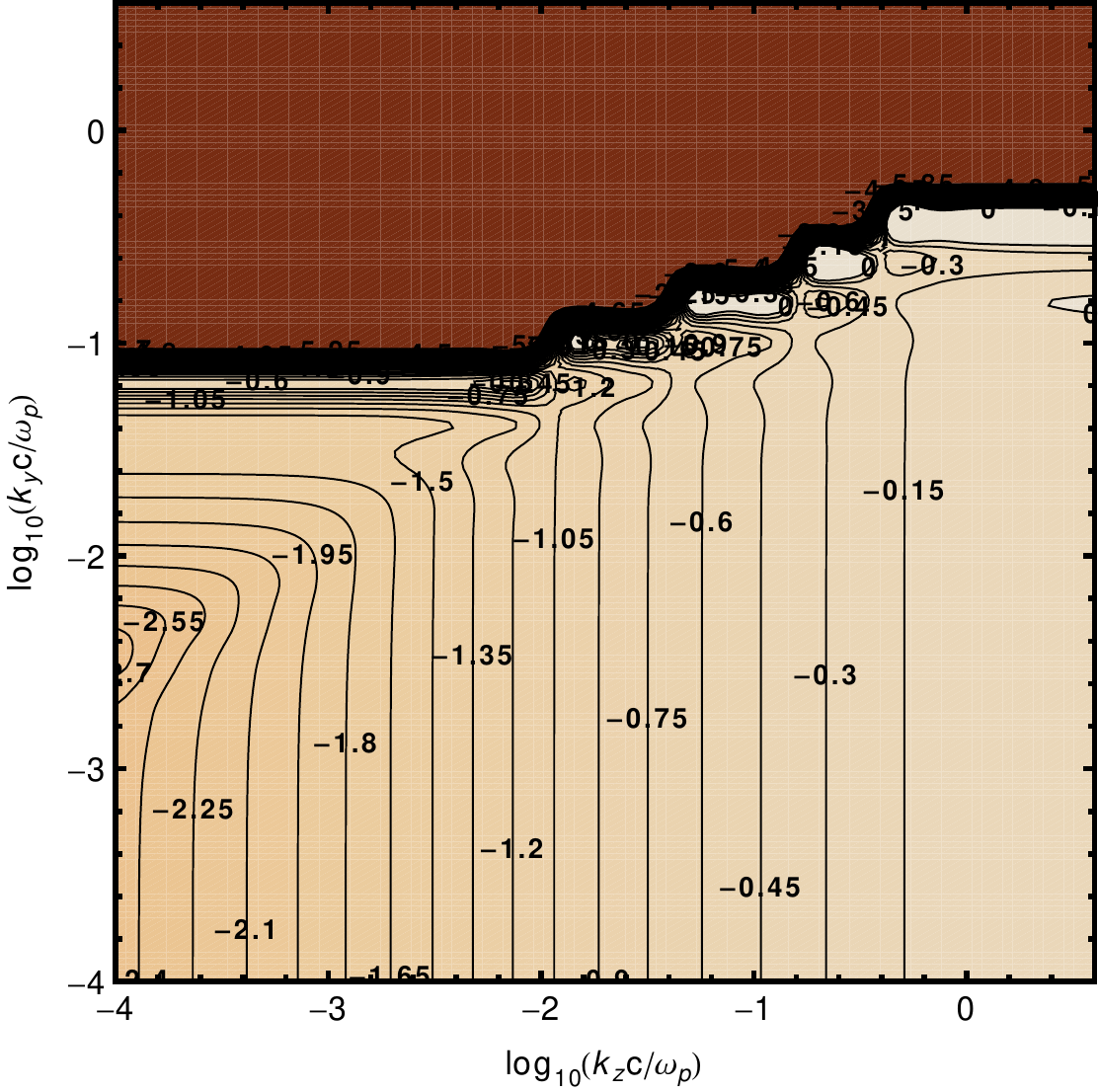}
\caption{Same as Fig.~\ref{fig:inst2dxz}, in the plane $(k_y,k_z)$,
  for $k_x=0$. The growth rate vanishes at large values of $k_y$.\label{fig:inst2dyz} }
\end{figure}

We now present numerical solutions of this dispersion relation in the
various 2D planes: $(k_x,k_z)$ in Fig.~\ref{fig:inst2dxz} assuming
$k_y=0$; $(k_x,k_y)$ in Fig.~\ref{fig:inst2dxy} assuming $k_z=0$; and
$(k_y,k_z)$ in Fig.~\ref{fig:inst2dyz} assuming $k_x=0$.

The global trend that emerges from these numerical simulations is,
here as well, a maximum growth rate of order $\beta_y\omega_{\rm p}$
at wavenumbers $\sim \omega_{\rm p}$, provided the thermal dispersion
velocity $\beta_{\rm s}$ remains much smaller than the drift velocity
$\beta_y$.

\subsection{Interpretation and analytical approximations}\label{sec:ana}
The above instability can be best understood in the limit $k_y\,=\,0$,
in the non-relativistic regime $\beta_y\,\ll\,1$, which formally
corresponds to $\gamma_{\rm sh}\xi_{\rm cr}\,\ll\,1$. In this limit,
one can neglect the acceleration of the plasma relative to the far
upstream, $\beta_{{\cal R}\vert\rm u}\,\sim\,0$, so that the
convective electric field can be neglected; furthermore,
$\kappa=\gamma_{{\cal R}\vert\rm u}/u^0\,\sim\,1$. Although
relativistic shock waves should rather lead to $\gamma_{\rm
  sh}\xi_{\rm cr}\,\gtrsim\,1$, we find little difference in the
growth rate between the above approximation and the numerical
calculation, suggesting that it remains a good approximation.

In this $\beta_y\,\ll\,1$ regime, the instability involves only
velocity fluctuations $\delta u_x$, $\delta u_z$, a density
fluctuation $\delta n$, and electromagnetic perturbations $\delta
B_x$, $\delta B_z$ and $\delta E_y$. One then finds that a combination
of the acoustic mode along $B_z$ and the high frequency superluminal
branch of the extraordinary mode is destabilized by the drift motion
that results from the compensation of the current $j_y$.

To see this, we use the perturbed $y-$component of the electromagnetic
vector potential $\delta A_y$ and the displacement $\boldsymbol{\xi}$
of the plasma. Maxwell equations then imply
\begin{equation}
\label{eq:dAy}
c^2 \boldsymbol{\nabla\cdot\nabla}\, \delta A_y -\partial_t^2\, \delta A_y + 4\pi c \delta j_y \,=\, 0,
\end{equation}
with $\delta j_y \,=\, \delta j_y^{\rm (r)} + \delta j_y^{\rm (c)}$,
with the following notations:
\begin{equation}
\delta j_y^{\rm (r)}\,\equiv\, 2e\,c\,n\,\Delta u_y
\end{equation}
and
\begin{equation}
\delta j_y^{\rm (c)}\,\equiv\,2e\,c\,\beta_y\,\delta n\ .
\end{equation}
Note the difference between $\Delta u_y$ and $\delta u_y$, defined in
Eq.~(\ref{eq:dn}). Note also that $u^0\,\sim\,1$ because we work here
in the rest frame of the ambient plasma under the approximation
$\gamma_{\rm sh}\xi_{\rm cr}\,\ll\,1$.

The response current $\delta j_y^{\rm (r)}$ evolves according to the
dynamical equation:
\begin{equation}
\label{eq:djr}
\partial_t \delta j_y^{\rm (r)} =\frac{\omega_{\rm p}^2}{4\pi}\left(\delta
  E_y - \delta u_x B_{\rm u}\right)
\end{equation}
The perturbed bulk velocity can be written: $\boldsymbol{\delta u}
= \partial_t \boldsymbol{\xi}$, and $\delta E_y\,=\, -\partial_t
\delta A_y/c$. Thus we obtain the simple relation
\begin{equation}
\label{eq:djr}
\delta j_y^{\rm (r)} = -\frac{\omega_{\rm p}^2}{4\pi c}\left(\delta A_y+B_{\rm u}\xi_x\right) \ .
\end{equation} 
The dynamics of the center of mass is governed by a MHD-type equation
(with $\rho = 2\,n\,m$):
\begin{equation}
\label{eq:dv}
\rho \partial_t \boldsymbol{\delta u} + \rho c_{\rm
  s}^2\boldsymbol{\nabla} \frac{\delta n}{n} 
\,=\, \frac{1}{c}\boldsymbol{j}\times \boldsymbol{\delta B} + 
\frac{1}{c}\boldsymbol{\delta  j^{\rm (r)}} \times \boldsymbol{B_{\rm u}} \ ,
\end{equation}
with of course, $\boldsymbol{j}\,\equiv\, 2 n e c u_y\,
\boldsymbol{y}$.  Note that $\delta j^{\rm (c)}$ does not contribute
to the Lorentz force because the term in $\delta n$ cancels out with
the equilibrium condition.  Note also that $\delta B_x = - \partial_z
\delta A_y$ and $\delta B_z = \partial_x \delta A_y$.  In particular
the $x-$component reads:
\begin{equation}
\label{eq:dvx}
\rho \partial_t^2 \xi_x + \rho c_{\rm s}^2 \partial_x \frac{\delta n}{n}
\,=\,\frac{1}{c}j_y \delta B_z + \frac{1}{c}\delta j_y^{\rm (r)}
B_{\rm u} \ ,
\end{equation}
which can be rewritten as (introducing $\delta\tilde
A_y\,\equiv\,\delta A_y/B_{\rm u}$):
\begin{equation}
\label{eq:dvx2}
\partial_t^2 \xi_x + c_{\rm s}^2 \partial_x \frac{\delta n}{n}  \,=\,
\omega_{\rm c} \beta_y c  \partial_x \delta \tilde A_y -
\omega_{\rm c}^2\left(\delta \tilde A_y + \xi_x\right) \ .
\end{equation}
One can use also the $z-$component, however it turns out that the
equation for sound evolution is more convenient; we obtain it by
taking the divergence of the dynamical equation:
\begin{equation}
\label{eq:dtn}
\left(\partial_t^2 - c_{\rm s}^2 \Delta\right) \frac{\delta n}{n} \,=\, 
\omega_{\rm c}^2 \partial_x \left(\delta \tilde A_y +
  \xi_x\right) - 
\omega_{\rm c}\beta_y c \Delta
\delta \tilde A_y \ .
\end{equation}
Therefore we have obtained three dynamical equations of second order
in time derivative that couple $\delta A_y$, $\delta n$ and
$\xi_x$. Equation~(\ref{eq:dAy}) for $\delta A_y$ can be rewritten as
\begin{equation}
  c^2\Delta \delta \tilde A_y - \partial_t^2 \delta \tilde A_y -
  \omega_{\rm p}^2 \delta \tilde A_y - \omega_{\rm p}^2\xi_x +
  \frac{\omega_{\rm p}^2}{\omega_{\rm c}}\beta_y \frac{\delta n}{n} \,=\,0\ .
\end{equation}

This system leads to the following dispersion relation:
\begin{eqnarray} 
\left[P_X(\omega^2) - \beta_y^2\omega_{\rm p}^2 k^2 c^2\right]
  \omega^2 + \beta_y^2\omega_{\rm p}^2 \omega_{\rm c}^2 k_z^2 c^2 && \nonumber\\
 \quad\quad-k^2 c_{\rm
    s}^2\left[P_X(\omega^2) + \frac{k_x^2}{k^2} \omega_{\rm c}^2
    (\omega^2-k^2c^2)\right] &\,=\,&0\label{eq:disp0}
\end{eqnarray}
with 
\begin{equation}
  P_X(\omega^2) \equiv \omega^4 -(\omega_{\rm p}^2 + \omega_{\rm c}^2
  +k^2 c^2) \omega^2 + \omega_{\rm c}^2 k^2c^2\ ,
\end{equation}
and $k^2=k_x^2+k_z^2$; $P_X(\omega^2) = 0$ gives the dispersion
relation of the extraordinary mode in the cold plasma limit.

This dispersion relation matches well Eq.~(\ref{eq:disp1}) up to
relativistic corrections in $\beta_y$. Let us discuss
Eq.~(\ref{eq:disp0}) in several limits of interest.

\subsubsection{Cold and weakly magnetized: $\beta_{\rm s}^2 \,\ll\,
  \sigma \,\ll\, 1$}
Let us analyze the instability in the cold plasma limit, and at small
values of $\sigma$, however not necessarily smaller than $\beta_y^2$
when this parameter is small. The dispersion relation reduces to:
\begin{equation}
\label{eq:disp0a}
P_X(\omega^2) - \omega_{\rm p}^2 k^2 \beta_y^2 c^2 = 0 \ .
\end{equation}
This leads to a negative root in $\omega^2$:
\begin{equation}
\label{eq:disp0b}
\omega_-^2 = - (\beta_y^2 - \sigma) \omega_{\rm p}^2 \, F(k^2 \delta^2) \ ,
\end{equation}
with $\delta \,\equiv\,c/\omega_{\rm p}$, and
\begin{equation}
\label{eq:disp0b1}
F(k^2 \delta^2) = \frac{2 k^2\delta^2}{1+k^2\delta^2  + 
\left[(1+k^2\delta^2)^2 + 4(\beta_y^2-\sigma)k^2\delta^2\right]^{1/2}} \ .
\end{equation}
In the latter expression, the contribution of $\sigma$ must be kept
when it is no longer negligible compared to $\beta_y^2$.  For $k^2
\delta^2 \ll 1$, $F(k^2 \delta^2) \simeq k^2 \delta^2$ and
\begin{equation}
\label{ }
\omega_-^2 \simeq - \omega_{\rm p}^2 (\beta_y^2 - \sigma)k^2 \delta^2 \ .
\end{equation}
For $k^2 \delta^2 \gg 1$, $F(k^2 \delta^2) \simeq 1$ and 
\begin{equation}
\label{ }
\omega_-^2 \simeq - \omega_{\rm p}^2 (\beta_y^2 - \sigma) \ ;
\end{equation}
which gives the maximum growth rate.  Clearly the instability occurs
at low magnetization, precisely when $\sigma < \beta_y^2$, in very
good agreement with the analysis of the previous Section.  

\subsubsection{Long wavelength modes, $k^2 \delta^2 \,\ll\, 1$ and finite $\sigma$}
In this limit $k^2 \delta^2 \,\ll\, 1$, we find
\begin{equation}
  \omega_-^2 \simeq - \beta_y \omega_{\rm p}^2 \sqrt{\frac{\sigma}{1+\sigma}} k_z \delta \ ;
\end{equation}
i.e. a growth rate for small $\sigma$
\begin{equation}
\Im\omega \,\simeq\, \sqrt{ \omega_{\rm c} k_z \beta_y c} \ ;
\end{equation}
which extends previous results obtained in Pelletier et al. (2009) and
in Casse et al. (2013) in the MHD regime for similar configurations;
see also Riquelme \& Spitkovsky (2010), Nekrasov (2013) for similar
configurations in the non-relativistic limit.  It thus indicates that
this instability has a kinetic origin and that the MHD solution
describes its long wavelength behaviour.

\subsubsection{Warm plasma with $\sigma \ll \beta_{\rm s}^2$}
From the general dispersion relation we find:
\begin{equation}
\omega_-^2 = -\omega_{\rm p}^2\left[\beta_y^2-\beta_{\rm s}^2(1+k^2 \delta^2)\right]
F_s(k^2 \delta^2, \beta_{\rm s}) \ ,
\end{equation}
where 
\begin{eqnarray}
F_s(k^2 \delta^2,\beta_{\rm s}) &\,= \,&
2 k^2\delta^2\biggl\{1+(1+\beta_s^2)k^2\delta^2 \nonumber\\
 &&+ \biggl[\left[1+(1+\beta_s^2)k^2\delta^2\right]^2 \nonumber\\
 &&+ 4(\beta_y^2-\beta_s^2(1+k^2\delta^2)k^2\delta^2)\biggr]^{1/2}\biggr\}^{-1} \ ,
\end{eqnarray}
which can be well approximated by
\begin{equation}
F_s(k^2 \delta^2) \,\simeq\, \frac{k^2\delta^2}{1+ k^2\delta^2} \ ,
\end{equation}
The main conclusion is that temperature effects quench the
instability when $\beta_{\rm s}\,\gtrsim\,\beta_y$, as reported in the
previous Section.

\subsection{Description and evolution}

The instability presents the character of a common current instability
that is triggered when the drift velocity is larger than the sound
velocity and also the character of a Weibel type electromagnetic
instability when the threshold is strongly overstepped. The growth
rate can reach values as large as $\omega_{\rm p}$ and makes the
instability faster than all instabilities previously studied,
including the filamentation instability triggered by the reflected
particles ($\Im\omega \,\simeq\, \sqrt{\xi_{\rm cr}} \omega_{\rm p}$),
the oblique two stream instability ($\Im\omega \,\simeq\, \xi_{\rm
  cr}^{1/3} \omega_{\rm p}$) etc.

We find that this instability is quenched at high temperatures, when
$\beta_y\,\lesssim\,\beta_{\rm s}\sim \sqrt{kT/mc^2}$. However, in the
precursor of relativistic shocks, one expects $\beta_y\,\sim\,1$ and
for pair shocks, the preheating inside the precursor remains
moderate. Therefore, such temperature effects are not expected to
contribute strongly.

\begin{figure}
\includegraphics[bb=60 150 680 520, width = 0.48\textwidth]{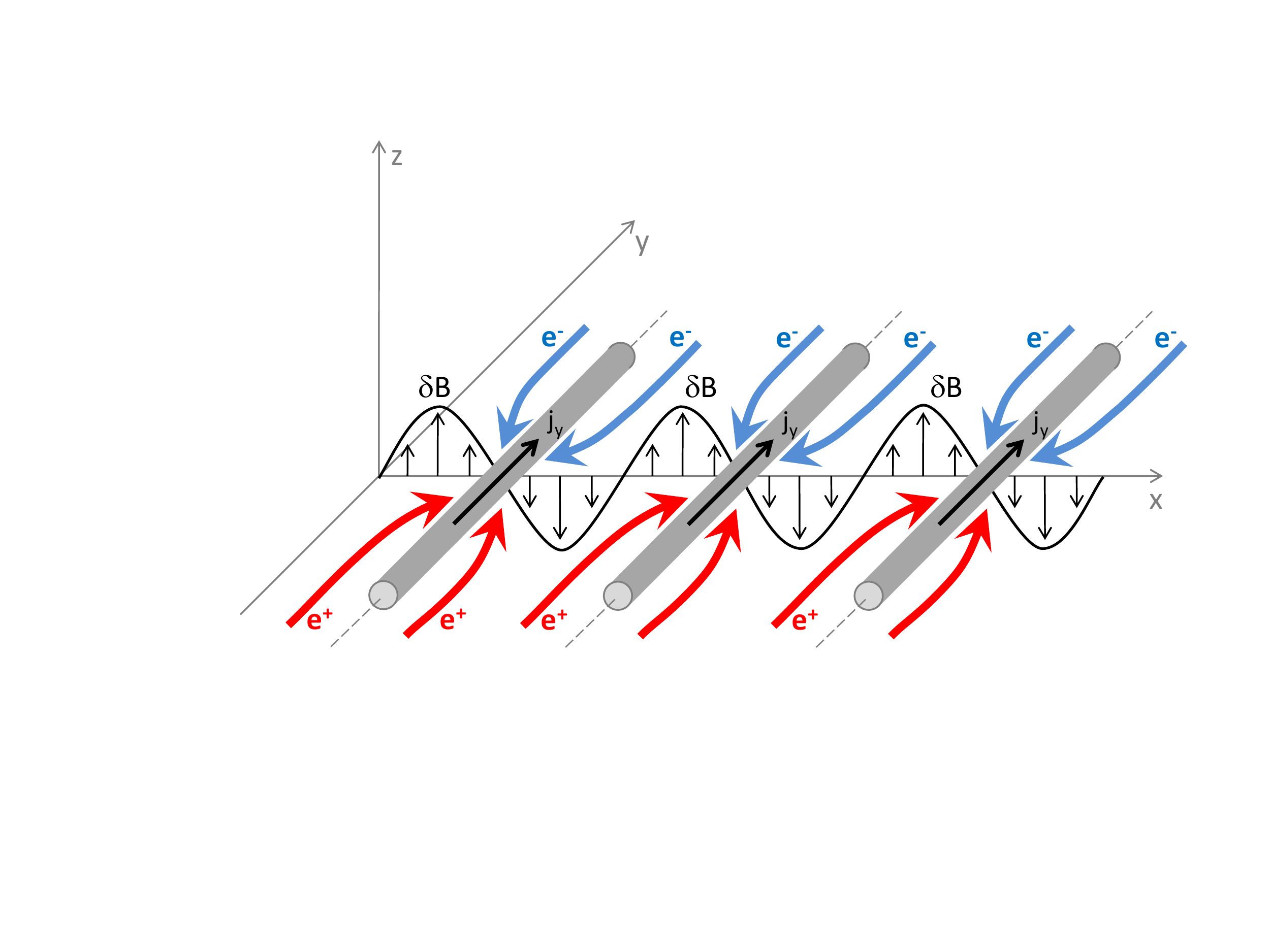}
\caption{Sketch of the development of the current-driven filamentation
  instability in 1D, with a perturbation along $x$. The current
  carried by the suprathermal particles, oriented along $-y$, is not
  indicated here. Electrons and positrons of the ambient plasma flow
  in opposite directions to compensate this current; if present, a
  magnetic fluctuation along $z$ leads to density enhancements along
  filaments, thereby creating a current perturbation which feeds back
  positively on the magnetic fluctuation. \label{fig:sk} }
\end{figure}

In the 2D setting $k_y=0$, this instability leads to filamentation of
the plasma in a way similar to the standard Weibel-filamentation
instability, with some noticeable differences. In particular, the
current perturbation is here produced by a global charge neutral
density variation, $\delta j_y^{\rm (c)} = \beta_y \delta n e c$, not
by a charge perturbation as in the Weibel/filamentation
instability. This density variation is itself produced by the
compression effect associated to the Lorentz force, derived from the
drift $\beta_y$. In contrast, the perturbed current in the
Weibel/filamentation instability $\delta j^{\rm (w)}\,=\, \beta_{\rm
  w} \delta \rho\, e c$, with $\beta_{\rm w}$ the drift velocity of
two $e^-$ counterstreaming beams (assuming that charge neutralization
is ensured, e.g. by ions) and $\delta \rho\, e$ the charge
perturbation (as before). The Lorentz force then couples this charge
perturbation to the electromagnetic potential through
\begin{equation}
  \partial_t^2 \delta \rho = -\beta_{\rm w} \frac{\omega_{\rm
      p}^2}{4\pi e c} \Delta \delta A_y \ .
\end{equation}
In this counterstreaming (symmetric) situation, the Weibel instability
gives rise to small scale magnetic perturbations with a growth rate
similar to that of the current-driven filamentation instability. The
difference pointed above, namely charge perturbation vs density
perturbation, brings in a major difference between these two
instabilities, which is related to the polarity of the current
filaments. While in the Weibel instability, the counterstreaming beams
contain particles of similar charge, which thus deviate in a perturbed
magnetic field in different directions to form filaments of opposite
current, in the current-driven filamentation instability, the beams
contain particles of opposite charge, which thus deviate in the same
direction and create filaments with a current oriented in the same
direction, i.e. so as to compensate the current of the suprathermal
particles. This picture is sketched in Fig.~\ref{fig:sk}.  Current
driven filamentation is thus subject to coalescence and
reconnection. The non-linear evolution of this instability will be
addressed in a forthcoming study (Plotnikov et al. 2013b).

\section{Discussion}\label{sec:disc}
In our treatment of the instability, we have neglected the response of
the plasma of suprathermal particles. This choice is dictated by
simplicity, as including the response involves doubling the number of
fluid variables, which renders the problem untractable. However, one
should expect this approximation to be valid at maximal growth rate,
since $\Im \omega$ then becomes larger than the plasma frequency of
the suprathermal particles, $\omega_{\rm pb} = \xi_{\rm cr}^{1/2}
\omega_{\rm p}$, with $\omega_{\rm p}$ the plasma frequency of the
ambient (upstream) plasma. As the instability develops
  and turbulence grows, one should of course expect the orbits of
  these suprathermal particles to deviate from their zeroth order form
  given in App.~\ref{sec:app}; this influence will be made more
  precise in the following Sec.~\ref{sec:disc2}.

\subsection{Relevance to relativistic shocks}\label{sec:cdir}
Let us now discuss why the current-driven filamentation instability is
likely to play a central role in shaping the precursor, the shock and
the acceleration process in the relativistic mildly magnetized regime.

Advection through the shock front provides a crucial limitation for
the growth of instabilities upstream of a relativistic shock front. In
the upstream plasma rest frame, this can be understood as follows: the
precursor extends at most to a distance $c/(\gamma_{\rm sh}\omega_{\rm
  c})$ (e.g. Milosavljevi\'c \& Nakar 2006, Pelletier et al. 2009),
because the suprathermal particles only rotate by an amount
$1/\gamma_{\rm sh}$ before being caught back by the shock front; this
takes a time $t_{\rm u}\,\sim\,\gamma_{\rm sh}\omega_{\rm c}^{-1}$,
but the distance between the shock front and the tip of this precursor
does not exceed $t_{\rm u}(1-\beta_{\rm sh}^2)\sim t_{\rm
  u}/(2\gamma_{\rm sh}^2)$. Therefore, as measured in the upstream
plasma rest frame (indicated by $_{\vert \rm u}$), any instability
whose growth rate $\Im\omega_{\vert{\rm u}}\,\lesssim\, \gamma_{\rm
  sh}\omega_{\rm c}$ cannot grow on the crossing time of the
precursor. For the filamentation instability, $\Im\omega_{\vert\rm
  u}\,\sim\, \xi_{\rm cr}^{1/2}\omega_{\rm p}$, therefore the
instability can grow only if $\gamma_{\rm sh}^2\sigma \xi_{\rm
  cr}^{-1}\,\lesssim\,1$ (Lemoine \& Pelletier 2010, 2011). This
indicates that mildly magnetized and/or large Lorentz factor shock
waves cannot be mediated by the Weibel-filamentation instability, as
mentioned in the introduction.

The present current-driven filamentation instability modifies this
picture, because it grows faster than any of the other instabilities
discussed in the context of relativistic shocks, and mostly because of
the impact of the current on the incoming plasma in the shock front
frame: as discussed in App.~\ref{sec:app} and Sec.~\ref{sec:linear},
if $\gamma_{\rm sh}\xi_{\rm cr}\,\gtrsim\,1$, the upstream cannot
compensate the current at rest; it is therefore accelerated along $x$
to a Lorentz factor $\gamma_{{\cal R}\vert\rm u}\,\sim\, \gamma_{\rm
  sh}\xi_{\rm cr}/2$ in the upstream rest frame, and its apparent
density increases by a similar amount. In the shock front frame, the
incoming plasma is slowed down to velocities $\beta_{x,\rm in}\,\sim\,
-(1-\xi_{\rm cr}^2/2)$, which means that the rest frame of the plasma
effectively moves with a Lorentz factor (along $\boldsymbol{-x}$):
$\gamma_{{\cal R}\vert\rm sh}\,\sim\,1/\xi_{\rm cr}$. This change of
rest frame, relative to far infinity, strongly modifies the criterion
under which the instability has or does not have time to grow. In the
${\cal R}$ frame, which defines the rest frame of the background
plasma after its acceleration phase, the shock front moves with a
Lorentz factor $\gamma_{{\cal R}\vert\rm sh}$, therefore the precursor
size extends to $c/\left(\gamma_{{\cal R}\vert\rm sh}\omega_{\rm
    c}\right)$ and the timescale for a plasma mode to cross this
precursor now reads
\begin{equation}
  t_{x\vert{\cal R}}\,\simeq\,\frac{1}{\gamma_{{\cal R}\vert\rm
      sh}\vert\beta_{{\cal R}\vert\rm sh}\vert \omega_{\rm c}}\ ,\label{eq:tR}
\end{equation}
so that the instability can grow whenever $\Im\omega\, t_{x\vert{\cal
    R}}\,\gtrsim\,1$, or
\begin{equation}
\sigma\,\lesssim\,\xi_{\rm cr}^2\ ,
\end{equation}
For typical values $\xi_{\rm cr}\,\sim\,0.1$, this implies that growth
is possible up to magnetization levels $\sigma \sim 10^{-2}$,
\emph{irrespective of the Lorentz factor of the shock}. The latter
point is of importance, because it guarantees the growth of
instabilities at large $\gamma_{\rm sh}$, for which the precursor
becomes very short in the upstream rest frame. This result appears
compatible with recent PIC simulations, as we argue in
Sec.~\ref{sec:pic}.

Once micro-turbulence grows upstream of a relativistic collisionless
shock, one may expect the Fermi process to develop (e.g. Lemoine et
al. 2006, Niemiec et al. 2006) although how well it develops depends
on the relative efficiency of scattering in the micro-turbulence
relatively to advection in the large scale field (Pelletier et
al. 2009, Lemoine \& Pelletier 2010). To discuss this on quantitative
grounds, we write the scattering frequency in the downstream rest
frame
\begin{equation}
\nu_{\rm s}\,\sim\, c \lambda_{\delta B}/r_{\rm
  g}^2\,\sim\, \epsilon_{B,\rm d} \left(\lambda_{\delta B}\omega_{\rm
    p}/c\right)\, \omega_{\rm p}\ ,
\end{equation}
$\epsilon_{B,\rm d}$ denoting an average value of the equipartition
fraction of the magnetic field downstream of the shock,
$\lambda_{\delta B}$ representing the coherence length of the field;
the above equation holds for typical supra-thermal particles of
Lorentz factor $\gamma_{\rm sh}$ in the downstream frame. As discussed
in Lemoine \& Pelletier (2010), scattering beats advection, hence the
Fermi process develops, when
\begin{equation}
  \nu_{\rm s}\,\gg\,\omega_{\rm c}\,\Leftrightarrow\,\sigma\,\ll\,\epsilon_{B,\rm
    d}^2\left(\lambda_{\delta B}\omega_{\rm
    p}/c\right)^{2} \ . \label{eq:nuo}
\end{equation}
PIC simulations suggest $\epsilon_{B,\rm d}\,\sim\,0.01$ and
$\lambda_{\delta B}\,\sim\,1-10c/\omega_{\rm p}$ with some degree of
uncertainty. Nevertheless, the above result indicates that the current
driven instability that we are discussing here must also play a key
role in the switch-on of the Fermi process, by building up the
micro-turbulence for any value of the shock Lorentz factor, up to
magnetization levels as high as $\sigma\,\lesssim\,10^{-2}$.

\subsection{Current-driven instability vs Weibel/filamentation}\label{sec:disc2}
At very low magnetization levels, one must expect this current-driven
filamentation to gradually disappear, once the other more standard
(Weibel-filamentation, two stream etc.)  instabilities can grow. To
see this, consider the extreme $\sigma\,\rightarrow\,0$ limit: the
Weibel/filamentation instability then grows, excites turbulence which
scatters the suprathermal particles; since this turbulence has no
preferred direction in the tranverse plane
$(\boldsymbol{y},\boldsymbol{z})$, no net perpendicular current arises
and current-driven filamentation does not take place.

At finite magnetization, the average current does not vanish, but it
may be randomized by the micro-turbulence. This effect has not been
taken into in the present calculations, which work at linear order and
which neglect the response of the cosmic rays.  In order to quantify
the magnitude of the back-reaction of the turbulence on the particle
trajectories, one must compare the upstream residence time derived
under the assumption that microturbulence controls the scattering
process with that derived assuming a coherent gyration in the
background field. Furthermore, this comparison must be made upstream,
in the proper frame of the micro-turbulence. In what follows, we
assume that this frame corresponds to ${\cal R}$.  In this ${\cal R}$
frame, the turbulent magnetic field strength $\delta B_{\cal
  R}\,\simeq\,\delta B/\gamma_{{\cal R}\vert\rm sh}$ relatively to
that measured in the shock front; similarly, the typical Lorentz
factor of a supra-thermal particle can be written
$\overline\gamma_{\cal R}\,\simeq\,\gamma_{{\cal R}\vert\rm
  sh}\gamma_{\rm sh}$; the background field has a strength
$B_{z\vert\cal R}\,\simeq\, \gamma_{{\cal R}\vert\rm u} B_{\rm u}$
[see Eq.~(\ref{eq:gru})]. In this ${\cal R}$ frame, return to the
shock takes place once the particle has been scattered by an angle
$\delta \theta_{\cal R}\,\sim\,1/\gamma_{{\cal R}\vert\rm sh}$ (see
the discussion in Milosavljevi\'c \& Nakar 2006, Plotnikov et
al. 2013a). If the supra-thermal particles gyrate coherently in the
background electromagnetic field, return occurs on a timescale
$t_{{\rm r}\omega\vert\cal R}\,\sim\,\delta \theta_{\cal
  R}/\omega_{\cal R}$, with $\omega_{\cal
  R}\,\simeq\,\left(\gamma_{{\cal R}\vert\rm u}/\overline\gamma_{\cal
  R}\right)\omega_{\rm c}$. If micro-turbulence controls the
scattering with scattering frequency $\nu_{\rm s\vert \cal
  R}\,\sim\,c\lambda_{\delta B\vert\cal R}/r_{\rm g\vert\cal R}^2$,
return takes place on a timescale $t_{{\rm r}\nu\vert\cal
  R}\,\sim\,\delta \theta_{\cal R}^2/\nu_{\rm s\vert\cal
  R}$. Comparing the two timescales leads to a critical magnetization
level:
\begin{equation}
\sigma_{\rm c}\,\sim\,\xi_{\rm cr}^2 \,\epsilon_{B,\rm
  u}^2\,\left(\lambda_{\delta B\vert\cal R}\omega_{\rm p}/c\right)^2\ .
\end{equation}
The quantity $\epsilon_{B,\rm u}$ denotes the typical level of
micro-turbulence, usptream of the shock, as measured in the shock
front frame. The factor $\xi_{\rm cr}$ appears in this formula because
the comparison has been made in the ${\cal R}$ frame.  If the upstream
magnetization $\sigma\,\lesssim\,\sigma_{\rm c}$, then micro-turbulent
scattering efficiently randomizes the trajectories in the shock front
plane, hence the perpendicular current as well. Conversely, if
$\sigma\,\gtrsim\,\sigma_{\rm c}$, the return trajectories maintain
their coherence, hence the current-driven instability develops
efficiently.

An interesting question is what happens at large Lorentz factors and
low magnetization levels $\sigma\,\ll\,\sigma_{\rm c}$, where feedback
from the turbulence should not be neglected, but where the
Weibel-filamentation instability does not have time to grow (in the
absence of slow-down of the plasma, see below). This area of parameter
space corresponds to $\sigma \gamma_{\rm sh}^2\xi_{\rm
  cr}^{-1}\,\gtrsim\,1$ and $\sigma \,\lesssim\,\sigma_{\rm c}$. Our
analysis suggests that the current-driven instability must develop at
the tip of the precursor, where the turbulence is sufficiently weak
that its back-reaction can be neglected. Furthermore, the deceleration
of the plasma, which results from current compensation, now allows the
Weibel/filamentation instability to grow: Eq.~(\ref{eq:tR}) indicates
that growth becomes possible in the ${\cal R}$ frame whenever
$\sigma\,\lesssim\,\xi_{\rm cr}^3$. This instability may then step
over closer to the shock front, where the back-reaction of the
turbulence strongly randomizes the return trajectories of the
supra-thermal particles.

Nevertheless, one expect the precursor to be shaped by the size
$c/\omega_{\rm c}$ if the current-driven instability shapes the
precursor, or even the tip of the precursor: beyond that length scale,
the turbulence must die away quickly, because the plasma has not yet
slowed down and instabilities cannot grow there; inside the precursor,
one may expect some form of equilibrium to be reached between the
level of the turbulence, the slow-down of the plasma and the growth
rate of the instabilities. Its detailed study lies beyond the present
work.

This description contrasts with what one expects in the region of
parameter space in which the Weibel-filamentation instability can grow
without the slow-down imposed by the current, i.e. $\sigma \gamma_{\rm
  sh}^2\xi_{\rm cr}^{-1}\,\lesssim\,1$ and
$\sigma\,\ll\,10^{-5}$. There, as discussed above, the current is
mostly randomized by the near isotropicity of the trajectories of
suprathermal particles in the shock front plane. In this limit, the
precursor extends to a scale $\epsilon_B^{-1}\left(\lambda_{\delta
  B}\omega_{\rm p}/c\right)^{-1}\,c/\omega_{\rm p}$, smaller than
$c/\omega_{\rm c}$, since the return of suprathermal particles is
controlled by the scattering in the small-scale turbulence
(Milosavljevi\'c \& Nakar 2006, Pelletier et al. 2009). This situation
actually matches the unmagnetized shock limit; hence, one may expect
to find a universal precursor profile, independent of the
magnetization parameter. The detailed discussion of the profile in
this regime is also left open for further study.

\subsection{Comparison to PIC simulations}\label{sec:pic}
Particle-in-cell simulations offer valuable tools to probe the
physics of relativistic collisionless shock waves. So far, most
studies have discussed the unmagnetized or strongly magnetized limit
and few have addressed the mild magnetization regime, of interest
here. We thus confront our findings to the recent simulations of
Sironi et al. (2013), which have explored the regime of moderate
magnetizations $\sigma \,=\,10^{-4} \rightarrow 10^{-2}$ at various
shock Lorentz factors $\gamma_{\rm sh}\,=\,5 \rightarrow 200$. Such
simulations are performed in the downstream plasma rest frame, which
does not differ much from the shock rest frame. In this rest frame,
the slow-down of the plasma along $\boldsymbol{x}$ is difficult to
measure, because the relative modification of $u_{x,\rm in}$ is only
of order $\xi_{\rm cr}$, see App.~\ref{sec:app}.

However, their Fig.~7 is particular interesting, because it reveals a
precursor whose profile does not depend on $\sigma$, provided one
rescales the distances by $\sigma^{1/2}\,=\,\omega_{\rm c}/\omega_{\rm
  p}$, i.e. provided the distances are expressed in units of
$c/\omega_{\rm c}$. It is actually possible to infer directly from
their figure the typical scale height of the precursor, $\sim
2c/\omega_{\rm c}$, with a rough exponential dependence. For the
parameters probed in this figure, $\gamma_{\rm sh}=21$
($=\sqrt{2}\gamma_0$ with their $\gamma_0=15$) and
$\sigma=10^{-4}\rightarrow 10^{-3}$, the Weibel-filamentation
instability cannot grow without the slow-down of the plasma imparted
by the current-driven filamentation. Therefore these simulations
directly probe the region of parameter space discussed above, in which
the current-driven filamentation instability plays the central
role. The structure of the precursor conforms well to the
expectations, with a size $\sim c/\omega_{\rm c}$.

In their Fig.~5, these authors show the magnetic structure of the
precursor in 3D simulations for similar parameters; the magnetic field
appears to be structured in sheets parallel to the $x-y$ plane rather
than filaments oriented along $\boldsymbol{x}$, which would be
expected for a standard Weibel/filamentation instability. Finally,
they report no dependence on the shock Lorentz factor, whereas a
rather strong dependence is expected if the Weibel-filamentation
instability alone shapes the precursor: as the line $\sigma
\gamma_{\rm sh}^2\xi_{\rm cr}^{-1}=1$ is crossed, one expects to
transit in a region in which the Weibel-filamentation instability can
no longer grow. This independence relative to the Lorentz factor
directly results from the slow-down imposed by the current
compensation in the $\gamma_{\rm sh}\xi_{\rm cr}\,\gg\,1$ limit:
inside the precursor, everything happens as if the shock were moving
relative to upstream with the Lorentz factor $\gamma_{{\cal R}\vert\rm
  sh}\,\sim\,1/\xi_{\rm cr}$, so that all memory of the initial
$\gamma_{\rm sh}$ is lost.

These trends strongly suggest that the present current-driven
filamentation instability shapes the precursor and the shock of weakly
magnetized ($\sigma\,\ll\,1$) relativistic shock waves.

Finally, the picture that we have elaborated in Sec.~\ref{sec:cdir}
also allows to understand, at least qualitatively, the results of
Sironi et al. (2013) concerning the development of Fermi
acceleration. Their simulations indicate that Fermi acceleration
develops for any value of the shock Lorentz factor, for magnetization
levels $\sigma\,\lesssim\,10^{-5}$. This conforms well with
Eq.~(\ref{eq:nuo}) and the discussion in Lemoine \& Pelletier
(2010). There, current-driven filamentation can grow, irrespectively
of the shock Lorentz factor; it builds up turbulence and, because
$\sigma\,\lesssim\,10^{-5}$, scattering in the micro-turbulence
downstream of the shock front beats advection, hence the Fermi process
develops. At larger values of $\sigma$, the same simulations indicate
that Fermi acceleration develops in a restricted dynamic range, with a
maximum energy scaling as $\sigma^{-1/4}$. Equation~(\ref{eq:nuo}),
taken at face value, would indicate that Fermi acceleration should not
develop in this limit. However, this argument assumes a homogeneous
micro-turbulence downstream of the shock, of strength $\epsilon_{B,\rm
  d}$, whereas the micro-turbulence seen in PIC simulations actually
decreases away from the shock front. If the law of evolution of
$\epsilon_B$ were known, one could improve on Eq.~(\ref{eq:nuo}) by
comparing the scattering time in this evolving micro-turbulence and
the gyration time in the background field. In the absence of such a
well-defined law, one can nevertheless understand on a qualitative
level the scaling of the maximal energy: as the magnetization
increases beyond $10^{-5}$, the condition
$\sigma\,\lesssim\,\epsilon_{B,\rm d}^2$ remains true only in a finite
layer close to the shock front; since the scattering length-scale
evolves as the square of the particle energy, the restricted size of
this layer leads to the existence of a maximal energy. Let us note,
that if this layer were of infinite extent, there would nevertheless
be a maximal energy, scaling as $\sigma^{-1/2}$, as dicussed in
Pelletier et al. (2009).

\subsection{Consequences}
The above discussion directly impacts our understanding of shock
structuration and of particle acceleration. For instance, Sironi et
al. (2013) argue that in front of the shock, there exists a layer of
size $\sim c/\omega_{\rm c}$ filled with Weibel turbulence at a level
$\epsilon_B \,\sim\,10^{-2}$; this observation is based on the
simulations reported above, in the range $\sigma =10^{-4}\rightarrow
10^{-2}$. According to the above discussion, this layer actually
reflects the constrained growth of current-driven filamentation and
Weibel- filamentation instabilities in the precursor, whose size is
set by the current profile, which extends on $c/\omega_{\rm c}$, and
the turbulence is not of Weibel origin. 

These authors then extrapolate their results to the regime of low
magnetization $\sigma \,\ll\,10^{-5}$ to discuss the maximal energy of
particles accelerated at relativistic shocks. The above arguments
indicate that such an extrapolation is not justified, because the
physics of the precursor are likely to change as one transits from the
region controlled by the current-driven filamentation instability to
that controlled solely by the Weibel-filamentation mode. In
particular, as $\sigma\rightarrow0$, the diverging scale
$c/\omega_{\rm c}$ must decouple and one expects the precursor profile
to be entirely controlled by the micro-turbulence, as in the
unmagnetized limit. The above discussion indicates that this
transition takes place close to the line $\sigma\gamma_{\rm
  sh}^2\xi_{\rm cr}^{-1}\,\sim\,1$ and $\sigma\,\sim\,10^{-5}$ to
$10^{-4}$; for $\gamma_{\rm sh}=21$ as used in these simulations, both
limits reduce to the latter $\sigma\,\sim\,10^{-5}$ to $10^{-4}$.

\section{Conclusions}\label{sec:conc}
This work reports on a new current-driven filamentation instability
usptream of a magnetized relativistic collisionless shock front. As
viewed in the shock front frame, the suprathermal particles, which are
reflected on the shock front, or accelerated at the shock, gyrate
around the perpendicular magnetic field in the shock precursor,
thereby depositing a strong current $j_{\rm cr}\,\sim\,\xi_{\rm
  cr}\gamma_{\rm sh}\, n_{\rm u}\,e\,c$, which is both perpendicular
to the magnetic field and to the shock normal. As the incoming plasma
enters the precursor, it seeks to compensate this current within a
few skin depth scales. If $\xi_{\rm cr}\gamma_{\rm sh}\,\gtrsim\,1$,
which is a likely situation for highly relativistic shocks, the
incoming plasma cannot compensate this current in the upstream rest
frame; it is thus accelerated to a large Lorentz factor $\sim \xi_{\rm
  cr}\gamma_{\rm sh}/2$ (relative to far upstream), which increases
the apparent density of the plasma by a similar factor; particles then
drift at relativistic velocities in the perpendicular direction to
achieve current compensation, electrons and positrons drifting in
opposite directions. In the shock front rest frame, the incoming
plasma is decelerated along the shock normal at the same time as it is
accelerated in this perpendicular direction.

As we have argued, this current destabilizes a combination of the high
frequency branch of the extraordinary mode and of the acoustic mode
along the magnetic field. In a 2D configuration, in which one neglects
perturbations along the direction of the current, this instability
bears some resemblance to the Weibel-filamentation
instability. However, in the present case, the electromagnetic
perturbation couples to a density fluctuation, not to a charge
fluctuation, because the counterstreaming electrons and positrons
carry opposite charges. This leads to the formation of current
filaments of a same polarity, all currents being oriented so as to
compensate the cosmic ray current induced in the precursor. We find
that this instability has a very fast growth rate, of order
$\Im\omega\,\sim\,\beta_y\,\omega_{\rm p}$ on skin depth scales, with
$\beta_y\,\sim\,1$ the drift velocity. This instability is likely to
play a key role in shaping the precursor of weakly magnetized
relativistic collisionless shocks, in which the growth of other
instabilities is very often impeded by the fast transit across the
precursor.

In particular, we have shown that this instability can grow at any
value of the Lorentz factor, provided the magnetization parameter
$\sigma \,\lesssim\,\xi_{\rm cr}^2\,\sim\, 10^{-2}$. The relative
independence to the Lorentz factor of the shock, which controls the
size of the precursor $c/(\gamma_{\rm sh}\omega_{\rm c})$ (upstream
rest frame), stems from the deceleration that the incoming plasma
suffers inside the precursor: the relative Lorentz factor between the
shock front frame and the rest frame of the plasma now falls to
$\gamma_{{\cal R}\vert\rm sh}\,\sim\,1/\xi_{\rm cr}$, independent of
$\gamma_{\rm sh}$. In this picture, the shock foot plays the role of a
buffer that transforms the interaction with the fast incoming flow
into a more moderate regime, depending on the parameter $\xi_{\rm
  cr}$, over a well defined distance $\xi_{\rm cr} c/\omega_{\rm c}$
(in the instantaneous rest frame of the incoming plasma).

In previous studies, we have argued that the filamentation, oblique
two stream modes etc., can grow only at small values of $\sigma$ and
moderate values of $\gamma_{\rm sh}$, e.g. such that $\sigma
\gamma_{\rm sh}^2\xi_{\rm cr}^{-1}\,\lesssim\,1$ for the
Weibel-filamentation mode (Lemoine \& Pelletier 2010,
2011). Otherwise, the incoming plasma transits faster across the
precursor than a growth time of the instability. Therefore, the
current-driven filamentation instability emerges as the leading
instability outside this region of parameter space. At very low
magnetizations, $\sigma\,\ll\,10^{-5}$, and in the region where the
standard filamentation mode can grow, the current-driven filamentation
instability should gradually disappear, as the turbulent small scale
electromagnetic fields randomize the return trajectories of the
suprathermal particles in the shock front plane. In this limit, one
transits to the unmagnetized limit, in which the precursor size is no
longer controlled by the background magnetic field, but by the profile
of the micro-turbulence.

Outside this region, up to $\sigma \,\sim\,10^{-2}$, the current
driven filamentation instability is likely to play a dominant role.
The interesting physics of the shock at low magnetizations and at
Lorentz factor so large that the standard Weibel-filamentation mode
cannot grow, deserves close scrutiny. In this region, the current
filamentation instability can grow in the absence of strong
microturbulence; however the very growth of this instability and of
the filamentation mode, thanks to the deceleration of the plasma,
builds up the small scale turbulence, which then back reacts on the
current profile. The profile of the precursor in this regime is left
open for further study.

Our analysis at linear level indicates that the growth rate of the
current-driven filamentation instability is maximal on plasma skin
depth scales. This does not affect previous results concerning the
maximal energy of accelerated particles, which assume micro-turbulence
set on skin depth scales, e.g. Kirk \& Reville (2010), Bykov et
al. (2012), Plotnikov et al. (2013a). 

\section*{Acknowledgments} 
This work has been financially supported by the \emph{Programme National
Hautes Energies (PNHE)}.

\appendix

\section{Profile of the precursor}\label{sec:app}
We construct here the profile of the precursor in the cold plasma
limit, in the shock rest frame. We seek here a 1D zeroth
  order stationary solution of the shock precursor, with
  $\partial_t\,=\partial_y\,=\,\partial_z\,=\,0$, as dictated by the
  geometry of the problem. The perturbation of this solution leads to
  the linear system discussed in Sec.~\ref{sec:sysa} and
  App.~\ref{sec:sys}. As discussed in Sec.~\ref{sec:linear}, the
  zeroth order solution is characterized by the profile of the
  magnetic field $\boldsymbol{B}\,=\,B_z\, \boldsymbol{z}$, the
  convective electric field $\boldsymbol{E}\,=\,E_y\,\boldsymbol{y}$,
  and the fluid four-velocities of the various species. 

Note that the $x-$component of the current density vanishes for both
incoming particles and for suprathermal particles, as a consequence of
the stationary state: current conservation $\partial_\mu
j^\mu_\alpha\,=\,0$ for any species $\alpha$ implies the conservation
law $\partial_x j^x_\alpha\,=\,0$; since the $x-$component of the
current density of incoming particles, summed over electrons and
positrons, vanishes as $x\rightarrow+\infty$, it also vanishes in the
precursor, and similarly for the suprathermal particles. As particles
gyrate in the $(\boldsymbol{x},\boldsymbol{y})$ plane, we set
$u_\alpha^z\,=\,0$, hence $j_{z,\alpha}\,=\,0$ for all species.

Furthermore, we do not expect any non-zero $E_x$ component to emerge
inside the precursor because of the charge symmetry of the pair
plasma. One can check that the above solution is self-consistent. In
particular, the magnetic field does not possess other components, as a
result of $\boldsymbol{\nabla}\cdot\boldsymbol{B}\,=0\,$,
$\partial_y\,=\,\partial_z\,=\,0$ and $j_z\,=\,0$. 

\subsection{Simplified MHD model}\label{sec:MHD}
In Sec.~\ref{sec:linear}, we provide a relativistic two-fluid
description of the instability, the term two-fluid referring to the
electrons and positrons of the incoming background plasma. This
description thus extends beyond any MHD picture of the instability, up
to the inertial scale of the pair plasma. Nevertheless, it is
instructive to describe briefly the structure of the precursor in an
ideal MHD picture, in which one assumes that the magnetic field
remains frozen in the plasma all throughout the precursor.

Treating the suprathermal particle component as a tenuous fluid
carrying a current density $\boldsymbol{j_{\rm cr}}\,=\,j_{y,\rm
  cr}\,\boldsymbol{y}$, with $j_{y,\rm cr}\,=\,-\gamma_{\rm
  sh}\xi_{\rm cr}n_{\rm u}e c$, the electric field is fixed through
the frozen-in condition:
\begin{equation}
E_y\,=\,\overline\beta_{x,\rm in}B_z\ ,\label{eq:ey}
\end{equation}
with $\overline\beta_{x,\rm in}$ denoting the center-of-mass
$3-$velocity $x-$component of the incoming plasma. Then
$\boldsymbol{\nabla\times E}\,=\,0$ imposes
$\partial_x\left(\overline\beta_{x,\rm in}B_z\right)\,=\,0$, or
\begin{equation}
\overline\beta_{x,\rm in}B_z\,=\, \beta_{\rm sh}B_{\infty}\ ,\label{eq:MHDBz}
\end{equation}
with $B_{\infty}\,=\,\gamma_{\rm sh}B_{\rm u}$.  To keep the analysis
brief, here, we assume $\gamma_{\rm sh}\xi_{\rm cr}\,\ll\,1$, meaning
that the velocity of the electrons/positrons of the background plasma
along the $\boldsymbol{y}$ direction is much smaller than $c$. This
allows to set $\overline \beta_{x,\rm in}\,\simeq\,\overline u_{x,\rm
  in}/\left(1+\overline u_{x,\rm in}^2\right)^{1/2}$ in the above
equations, with $\overline u_{x,\rm in}$ the $x-$component of the
center-of-mass $4-$velocity.

The current density flowing in the incoming background plasma is
itself fixed through
\begin{equation}
j_{y,\rm in}\,=\,-j_{y,\rm cr} - \frac{c}{4\pi}\partial_x B_z\ .\label{eq:MHDjy}
\end{equation}

Particle number conservation $\partial_x\left(n_{\rm u}
u_x\right)\,=\,0$ and energy-momentum conservation in the cold plasma
limit then lead to the equation:
\begin{equation}
n_{\rm u}m_e c^2 \overline u_{x,\rm in}\partial_x \overline u_{x,\rm
  in}\,=\,\frac{1}{c}j_{y,\rm in}B_z\ .
\end{equation}
This equation of motion becomes an equation for $\overline u_{x,\rm
  in}$, once Eqs.~(\ref{eq:MHDBz}) and (\ref{eq:MHDjy}) have been taken
into account. This equation can be rewritten in the following compact
form:
\begin{equation}
\left[1 - \frac{\beta_{\rm sh}^2\gamma_{\rm sh}^2\overline\beta_{x,\rm
      in}^2}{\overline u_{x,\rm in}^4}\sigma\right]\,\overline
\beta_{x,\rm in}\overline u_{x,\rm in}\partial_x \overline u_{x,\rm
  in} \,=\,\beta_{\rm sh}\gamma_{\rm sh}\xi_{\rm cr}\frac{\omega_{\rm
    c}}{c}\ .\label{eq:dbx}
\end{equation}
In this equation, we have used the definition of the magnetization
parameter, Eq.~(\ref{eq:sigma}) and $\omega_{\rm c}\,=\, e B_{\rm
  u}/(m_e c)\,=\, e B_{\infty}/\left(\gamma_{\rm sh}m_e c\right)$.
Equation~(\ref{eq:dbx}) is particularly useful, because it allows to
obtain a quick estimate of the slow-down of the plasma due to the
Lorentz force: one first notes that the second term in the brackets,
which originates from the uncompensated part of the current in the
precursor, is much smaller than unity, and can be safely neglected;
then, one finds that between entry into the precursor and shock
crossing, the variation of $\overline u_{x,\rm in}$ reads
\begin{equation}
\Delta \overline u_{x,\rm in}\,\simeq\,-u_\infty\xi_{\rm
  cr}\ ,\label{eq:uxMHD}
\end{equation}
with $u_\infty\,=\,\gamma_{\rm sh}\beta_{\rm sh}$; note that the scale
of variation is set by the precursor size $c/\omega_{\rm c}$. Assuming
now that the transverse $3-$velocity of electrons and positrons, along
$\boldsymbol{y}$, is of order $\pm \gamma_{\rm sh}\xi_{\rm cr}$, one
can check that the $3-$velocity $\overline \beta_{x,\rm in,0}$ close
to the shock front is of the order of
\begin{equation}
\overline \beta_{x,\rm in,0}\,\simeq\, \beta_{\rm sh}\,\left(1 -
\frac{\xi_{\rm cr}^2}{2}\right) \ .
\end{equation}
These results will remain true in the following multi-fluid description,
even at large values of the quantity $\gamma_{\rm sh}\xi_{\rm cr}$. In
the above MHD model, $\partial_x B_z/B_z\,=\,- \partial_x \overline
\beta_{x,\rm in}/\overline\beta_{x,\rm in}$, therefore the above
scalings allow to derive an estimate of $(c/4\pi) \partial_x B_z$,
which characterizes the departure from current compensation in the
precursor:
\begin{equation}
\left\vert\frac{4\pi j_{y,\rm cr}}{c\,\partial_x B_z}\right\vert\,\approx\,
\frac{1}{\xi_{\rm cr}\sigma}\,\gg\,1\ ,
\end{equation}
indicating that the current is indeed compensated to very high
accuracy in the precursor.

\subsection{Multi-fluid model}\label{sec:coldprec}
We now turn to a more exhaustive multi-fluid model of the precursor,
which is necessary to construct the steady state on which the linear
analysis of Sec.~\ref{sec:linear} relies. In particular,
  we relax the frozen-in condition of the magnetic field inside the
  precursor and we follow the kinematics of the various particle
  populations along the $\boldsymbol{y}$ direction. Of course, well
  outside the precursor, one still assumes $E_y\,=\,\beta_{\rm
    sh}B_z$, corresponding to the assumption of zero electric field in
  the rest frame of the background plasma as $x\rightarrow +\infty$.

We consider the following populations of particles: the incoming
particles, denoted by the subscript $_{\rm in}$, and the suprathermal
particle population, which we divide into two sub-populations, those
moving toward $\boldsymbol{+x}$ from the shock front up to the tip of
the precursor (subscript $_{\rm r+}$) and those moving toward
$\boldsymbol{-x}$ from the tip of the precursor toward the shock front
(subscript $_{\rm r-}$). We set the shock front at $x=0$ and the tip
of the precursor at $x_1$. All throughout this section,
  we denote by $u^\mu_\alpha$ the four-velocity of the \emph{positron
    component} of species $\alpha$, with $\alpha\,\in\,\left({\rm
    in},{\rm r+},{\rm r-}\right)$. As discussed in
  Sec.~\ref{sec:linear}, the $x-$ components of the $4-$velocities of
  the electrons match those of the positrons, while the $y-$components
  are opposite.

Alsop \& Arons (1988) have described the structure of the precursor of
a strongly magnetized relativistic shock; they do so by solving the
fluid and Maxwell equations with one population of incoming particles,
which gyrate in the compressed magnetic field. The present description
is slightly different: we set a boundary at $x=0$, corresponding to
the shock transition, into which the incoming population flows and out
of which the suprathermal particle population emerges, with no
specific relation between these two populations.

In the cold plasma limit, the coherent rotation of the suprathermal
particles at the tip of the precursor implies $\beta_{x,\rm
  r\pm}(x_1)=0$, therefore $n_{\rm r\pm}(x_1)\,\rightarrow\,+\infty$,
and consequently $\vert j_y(x_1)\vert \,\rightarrow\,\infty$. This
singular behaviour disappears of course when warm plasma effects are
introduced.  Indeed, the suprathermal particle population should be
described in the present shock rest frame as a relativistically hot
plasma with mean Lorentz factor $\sim\gamma_{\rm sh}$ and roughly
isotropic distribution function. Such effects are discussed in the
next App.~\ref{sec:app-warm}. The cold plasma approximation, which we
use here, has the advantage of providing quantitative estimates for
the various quantities used in the manuscript.

The electromagnetic profile is thus determined by
$B_z\,\equiv\,\gamma_{\rm sh}B_{\rm u}(1+b)$, by the current $j_y$ and
the four-velocities of the respective fluids. This profile of the
precursor can be solved as a shooting problem, with three parameters
to be determined by the boundary conditions: $b_1$, $\gamma_{\rm r1}$,
corresponding respectively to the deviation from $\gamma_{\rm
  sh}B_{\rm u}$, the Lorentz factor of suprathermal particles at the
tip of the precursor, and $x_1$. The boundary conditions are:
\begin{eqnarray}
  u_{x,\rm r+,0}\,=\,u_{\rm sh}\ ,\quad
  u_{y,\rm r+,0}\,=\,0 \ , \nonumber\\
  n_{\rm r-,0}u_{y,\rm r-,0} \,=\,- n_{\rm in,0}u_{y,\rm in,0}\ .
\end{eqnarray}
The first two conditions specify the inital data for the suprathermal
particle population: we have chosen here a normal incidence to the
shock front and a Lorentz factor $\gamma_{\rm sh}$, as expected at
relativistic shocks. The third condition imposes a vanishing net flux
of particles along the shock front in the $\boldsymbol{y}$ direction.

In the cold plasma limit, and under the stationary state approximation
$\partial_t\,=\,0$, the fluid equations 
  $\partial_\mu\left(n_\alpha u_\alpha^\mu\right)\,=\,0$ and
  $\partial_\mu T^{\mu\nu}_{\alpha}\,=\,+e\, n_\alpha u^\mu_\alpha
  F^{\nu}_{\,\,\mu}$ (for the positron components) read:
\begin{eqnarray}
\partial_x\left(n_\alpha
u_\alpha^x\right)&\,=\,&0\ ,\nonumber\\ 
\beta_{x,\alpha}\,\partial_x u^x_{\alpha}&\,=\,&\frac{e}{m_e}\beta_{\alpha,y}B_z\ ,\nonumber\\ 
\beta_{x,\alpha}\,\partial_x u^y_{\alpha}&\,=\,&\frac{e}{m_e}\left(E_y
- \beta_{\alpha,x}B_z\right)\ ,\nonumber\\
\beta_{x,\alpha}\,\partial_x
\gamma_\alpha&\,=\,&\frac{e}{m_e}\beta_{y,\alpha}E_y\ .\label{eq:prec0}
\end{eqnarray}
Here, $\gamma_\alpha\,\equiv\,u^0_\alpha$.  For the various species,
the continuity equations imply that at each point: $n_{\rm in}u_x
\,=\, n_{\infty}u_{\infty}$ with $u_{\infty}=\gamma_{\rm sh}\beta_{\rm
  sh}<0$, $n_{\rm r+}u_{x,\rm r+}\,=\,\xi_{\rm cr} n_{\infty}u_{\rm
  x,r+,0}$, $n_{\rm r-}u_{x,\rm r-}\,=\,-\xi_{\rm cr} n_{\infty}u_{\rm
  x,r+,0}$. The quantity $n_\infty$ represents the proper particle
density as $x\rightarrow +\infty$, while $u_{x,r\pm,0}$ represents the
$x-$component of the 4-velocity of species $r\pm$ at the shock front.

Complemented with Amp\`ere's law $\partial_xB_z\,=\,-4\pi j_y/c$, the
system Eq.~(\ref{eq:prec0}) may then be rewritten:
\begin{eqnarray}
  \beta_{x,\alpha}\partial_x\beta_{x,\alpha} &\,=\,& \frac{\omega_{\rm L,\alpha}}{c}\left[(1+b)-\beta_{x,\alpha}\beta_{\rm
      sh}\right]\beta_{y,\alpha}\ ,\nonumber\\
  \beta_{x,\alpha}\partial_x \beta_{y,\alpha} &\,=\,& \frac{\omega_{\rm L,\alpha}}{c}\left[\beta_{\rm
      sh}\left(1-\beta_{y,\alpha}^2\right)-\beta_{x,\alpha}(1+b)\right]\ ,\nonumber\\
  \beta_{x,\alpha}\partial_x\gamma&\,=\,& \frac{\omega_{\rm L,\alpha}}{c}\gamma\beta_{y,\alpha}\beta_{\rm sh}\ ,\nonumber
  \\
\partial_x b&\,=\,& -\frac{\omega_{\rm c}}{\sigma \gamma_{\rm sh}
  n_\infty c}
\left(n_{\rm in}u_{y,\rm in} + n_{\rm r+}u_{y,\rm r+}
  + n_{\rm r-}u_{y,\rm r-}\right) \ ,\nonumber\\
&& \label{eq:prec}
\end{eqnarray}
with $\omega_{\rm L,\alpha}\,\equiv\,e\gamma_{\rm sh}B_{\rm
  u}/(\gamma_\alpha mc)\,=\,\omega_{\rm c}\gamma_{\rm
  sh}/\gamma_\alpha$, in terms of $\omega_{\rm c}\,\equiv\,eB_{\rm
  u}/(mc)$ the upstream cyclotron frequency, which sets the spatial
scale $c/\omega_{\rm c}$ of the precursor. As discussed above,
$u_{y,\rm in}$, $u_{y,\rm r+}$ and $u_{y,\rm r-}$ represent the
$y-$components of the $4-$velocities of the incoming, suprathermal
${\rm r+}$ and ${\rm r-}$ positron components respectively. The last
equation for $b$ implicitly uses the fact that the $y-$velocities of
electrons are opposite to those of the positrons, for both incoming
and suprathermal particles, hence their $y-$current densities add up;
the magnetization $\sigma$ is defined in Eq.~(\ref{eq:sigma}).  This
last equation holds in the shock precursor where the various
populations mix.

Given the above three parameters $b_1$, $\gamma_{\rm r1}$ and $x_1$,
these fluid equations must then be matched to the boundary conditions;
this determines the profile of the precursor.

\begin{figure}
  \includegraphics[bb= -15 0 530 400, width=0.49\textwidth]{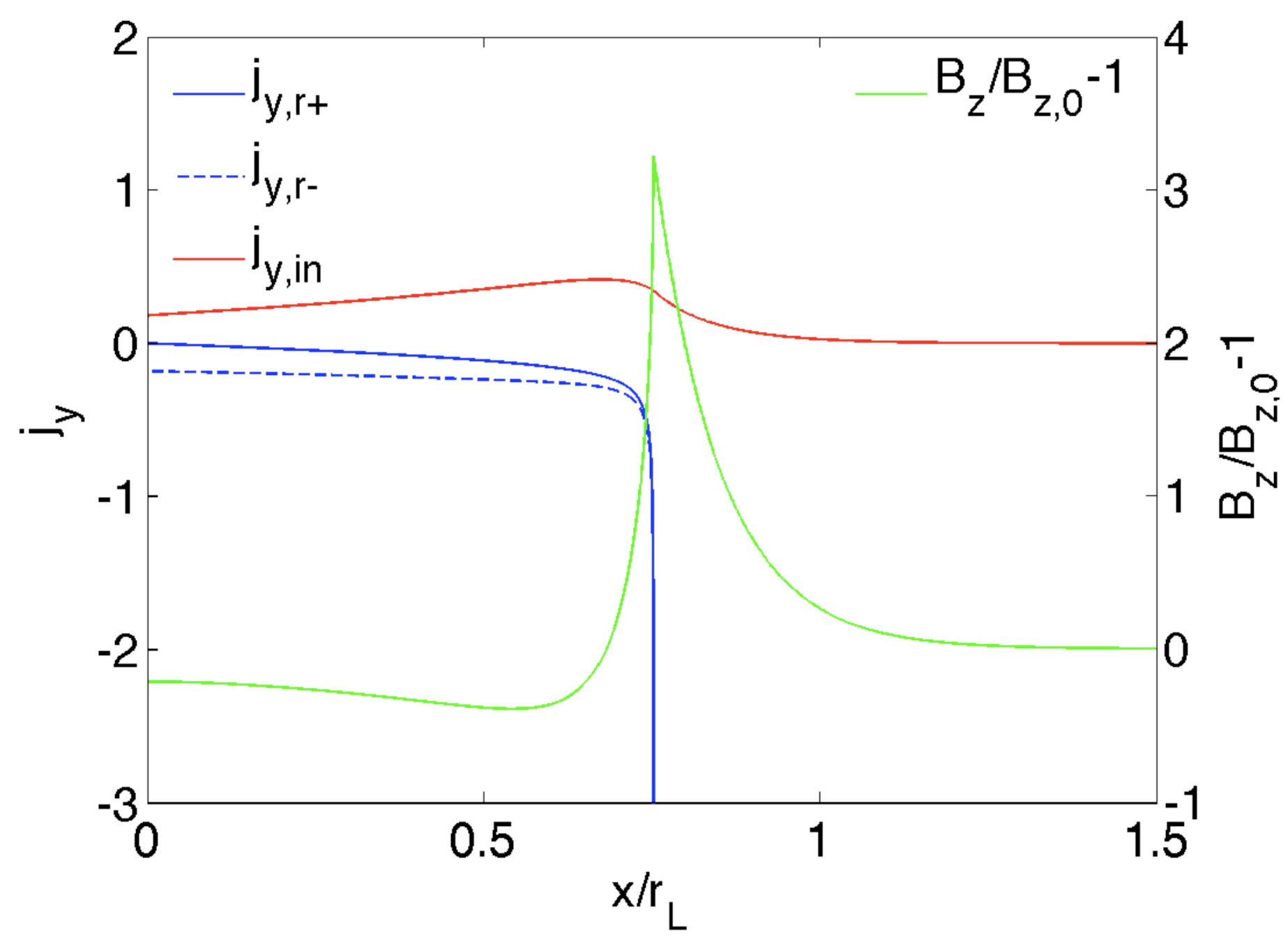}
  \includegraphics[bb= 0 0 530 400, width=0.49\textwidth]{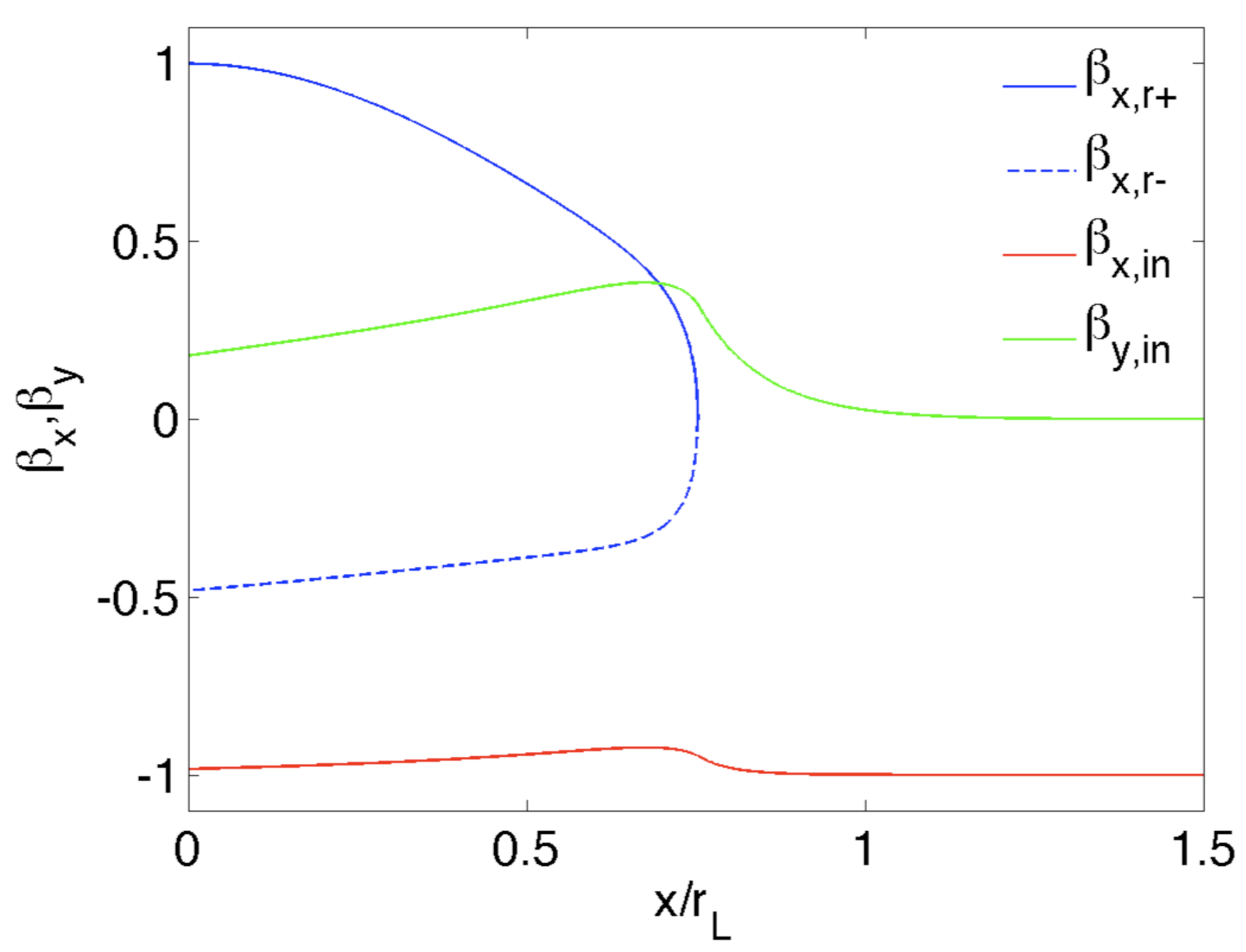}
  \caption{Structure of the precursor for $\sigma=0.01$, $\xi_{\rm
      cr}=0.1$ and $\gamma_{\rm sh}=1000$ ($r_{\rm
      L}\,\equiv\,c/\omega_{\rm c}$). Top panel: spatial profiles of
    the $y-$current carried by the suprathermal particle population in
    units of $e n_{\rm u}c$ (in blue), and of the compensating current
    carried by the inflowing ambient plasma (in red); in green, the
    spatial profile of the perturbed magnetic field
    $b=B_{z}/B_{z,0}-1$.  Bottom panel: spatial profiles of the
    $x-$velocities of the suprathermal particle population (in blue),
    of the inflowing background plasma (in red), and of the
    $y-$velocity of the background plasma positrons (in
    green).  \label{fig:prec} }
\end{figure}

Numerical examples of the profile are represented in
Fig.~\ref{fig:prec}. We have set $\sigma=0.01$, $\xi_{\rm cr}=0.1$ and
$\gamma_{\rm sh}=10^3$, but the profile does not depend on
$\gamma_{\rm sh}$ in the ultra-relativistic limit; it is entirely
controlled by $\sigma$ and $\xi_{\rm cr}$.

One can obtain an approximation to the above profile as follows. In
the vicinity of $x_1$, $n_{\rm r\pm}(x)\,\gg\,n_{\rm in}(x)$,
therefore the incoming particle contribution to Amp\`ere's law can be
neglected. Furthermore, one can approximate the motion of r+ particles
close to $x_1$ as uniform deceleration, implying
$\beta_x\,\simeq\,\left[2\vert\dot\beta_x(x_1)\vert(x-x_1)\right]^{1/2}$,
with $\vert\dot\beta_{x}(x_1)\vert\,=\,\omega_{\rm
  L1}(1+b_1)\beta_{y,\rm r+}(x_1)$ given that $\beta_x(x_1)=0$, and
$\omega_{\rm L1}\,\equiv\,\omega_{\rm L,r+}(x_1)$. This allows to
determine the singular profile of the density close to $x_1$, using
the continuity equation. Plugging this result and the similar estimate
for r- particles into Amp\`ere's law, one derives
\begin{equation}
  b\,\simeq\, b_1\,\left[1 + \frac{\xi_{\rm
        cr}}{\sigma}\frac{\vert\beta_{x,\rm
        r+}(0)\vert\sqrt{\beta_{y,\rm
          r+}(x_1)}}{\sqrt{2}(1+b_1)^{1/2}b_1}
    \omega_{\rm L1}^{1/2}\left(x-x_1\right)^{1/2}\right]\ .\label{eq:b0}
\end{equation}
The term in the brackets determine the scale over which $b$ varies
close to $x_1$, $\Delta x \,\sim\, \sigma^2\xi_{\rm
  cr}^{-2}b_1^3c/(\sqrt{2}\omega_{\rm L1})$. Using Amp\`ere's law with
$\partial_x b\,\sim\,b_1/\Delta x$, $\omega_{\rm
  L1}\,\sim\,\omega_{\rm c}$ and assuming $b\,\gg\,1$ leads to
\begin{equation}
  b_1 \,\sim\, \left(\frac{\xi_{\rm
        cr}}{\sigma}\right)^{1/2}\ .\label{eq:b1}
\end{equation}
The above turns out to provide the correct scaling seen in the
numerical calculations. In turn, this leads to $\Delta x \sim
\sigma^{1/2}\xi_{\rm cr}^{-1/2}c/\omega_{\rm c}\,\sim\,\xi_{\rm
  cr}^{-1/2}c/\omega_{\rm p}$: current compensation takes place on
skin depth scales, as anticipated in Lemoine \& Pelletier (2011).

Outside the precursor, the field goes down to its asymptotic far
upstream value on skin depth scales as well. Equations~(\ref{eq:prec})
can be used in this region, with $n_{\rm r\pm}\rightarrow 0$ in
Amp\`ere's law. As discussed in Alsop \& Arons (1988), the system
then admits the two integrals of motion
\begin{eqnarray}
\gamma_{\rm in}&\,=\,&\gamma_{\rm sh}\left(1-\sigma b\right)\ ,\nonumber\\
u_{x,\rm in}&\,=\,& u_\infty\left[1-\frac{\sigma}{2\beta_{\rm
      sh}^2}b\left(b+2\right)\right]\ .
\end{eqnarray}
These two integrals, combined with Eqs.~\ref{eq:prec} allow to derive
the following equation for the profile of $b$:
\begin{equation}
\partial_xb\,=\,-\frac{\omega_{\rm c}}{\sqrt{\sigma}c}\frac{\left[b^2 -
      \sigma b^2/(\gamma_{\rm sh}^2\beta_{\rm sh}^2) - \sigma
      b^3/\beta_{\rm sh}^2 - \sigma b^4/(4\beta_{\rm
        sh}^2)\right]^{1/2}}{1 - \sigma b(b+2)/(2\beta_{\rm sh}^2)}\ .
\end{equation}
This equation reveals the length scale of the profile: $c/\omega_{\rm
  p}$, and allows to solve for $b$, by integrating from $b_1$ up to
$+\infty$, then for $u_{\rm in}$. 

Using the integrals of motion, one computes the typical change in
Lorentz factor at the entrance into the precursor, 
\begin{eqnarray}
  \gamma(x_1)
  &\,=\,& \gamma_{\rm sh}\left(1 - \sqrt{\sigma\xi_{\rm cr}}\right)\ ,\nonumber\\
  u_{x,\rm in}(x_1)&\,\simeq\,& u_{\infty}\left(1-\xi_{\rm
    cr}/2\right) \ ,\label{eq:ux0}\\
  \vert u_{y,\rm in}(x_1)\vert &\,\simeq\,& -u_{\infty}\sqrt{\xi_{\rm cr}}\label{eq:uy0}\ .
\end{eqnarray}
The variation in Lorentz factor is small compared to that of $u_x$ and
$u_y$, but the slow-down along $x$ is substantial: at $x_1$, the
particles move at velocity $\beta_{x,\rm in}(x_1)\,\simeq\, 1-\xi_{\rm
  cr}/2$ in the shock front frame. This slow-down is obvious in
Fig.~\ref{fig:prec}.

Well inside the precursor, current compensation implies 
\begin{equation}
\vert u_{y,\rm  in}\vert \,\simeq\, \xi_{\rm cr}\gamma_{\rm sh}\ .\label{eq:uyf}
\end{equation}
In order to derive the slow-down imparted to incoming particles, one
first notes that $b\,\ll\,1$ outside the peak at the tip of the
precursor, as indicated by Eqs.~(\ref{eq:b0}) and (\ref{eq:b1}).  The
dynamics of incoming particles is then given by Eq.~(\ref{eq:prec})
with $b\,\ll\,1$, which implies that the flow is slowed by an amount
\begin{equation}
\Delta u_{x,\rm in} \,\simeq\,\gamma_{\rm sh} \xi_{\rm cr}\ ,\label{eq:uxf}
\end{equation}
between the far upstream value and the value of $u_{x,\rm in}$ well
inside the precursor. This value matches that at entry into the
precursor, Eq.~(\ref{eq:ux0}), and it also matches the value obtained
in the simplified MHD model, Eq.~(\ref{eq:uxMHD}). This slow-down
appears as a direct consequence of current compensation, which imposes
a Lorentz force directed in the $\boldsymbol{+x}$ direction.  In a
similar way, one derives $\Delta\gamma_{\rm in}\,\sim\,-\gamma_{\rm
  sh}\xi_{\rm cr}$. Thus the Lorentz factor of both flows remains
large after its modification by the Lorentz force. In terms of
3-velocity, this implies that $\boldsymbol{\beta}^2$ remains close to
unity, up to $1/(2\gamma_{\rm sh}^2)$. Using Eqs.~(\ref{eq:uyf}) and
(\ref{eq:uxf}), one derives the $3-$velocities well inside the
precursor: 
\begin{equation}
\vert\beta_y\vert \,\simeq\, \xi_{cr}\ ,
\end{equation}
and, at large values of $\gamma_{\rm sh}$,
\begin{equation}
\beta_{x,\rm in}\,\simeq\, \beta_{\rm sh}\left(1-\xi_{\rm cr}^2/2\right)\ .
\end{equation}
Therefore, if $\gamma_{\rm sh}\xi_{\rm cr}\,\gg\,1$, then
$\beta_{x,\rm in}\,\simeq\, - \left(1-\xi_{\rm cr}^2/2\right)$, while
$\beta_{x,\rm in}\,\simeq\,\beta_{\rm sh}$ in the opposite limit,
which corresponds to negligible, sub-relativistic deceleration.

Assuming that $\gamma_{\rm sh}\xi_{\rm cr}\, \gg\,1$, the relative
Lorentz factor between the shock front frame and the frame ${\cal R}$
in which the incoming is at rest along $\boldsymbol{x}$,
i.e. $u_{x,\rm in\vert{\cal R}}\,\equiv\,0$, has fallen from $\gamma_{\rm
  sh}$ outside the precursor down to
\begin{equation}
  \gamma_{{\cal R}\vert\rm sh}\,\simeq\, \frac{1}{\xi_{\rm cr}}\ .\label{eq:gin}
\end{equation}
As viewed in the upstream rest frame, the ambient plasma has been
picked up by the current layer and accelerated towards
$\boldsymbol{+x}$ to a Lorentz factor
\begin{equation}
  \gamma_{{\cal R}\vert\rm u}\,\simeq\, \gamma_{\rm sh}\frac{\xi_{\rm
      cr}}{2}\quad\quad \left(\gamma_{\rm sh}\xi_{\rm
      cr}\,\gg\,1\right)\ .
\end{equation}

Finally, in the ${\cal R}$ frame in which the ambient plasma is at
rest, the particles move with velocity $\vert\beta_{y,{\rm
    in}\vert{\cal R}}\vert\,\sim\,1$ with bulk Lorentz factor $\sim
\gamma_{\rm sh}\xi_{\rm cr}/2$, provided of course that $\gamma_{\rm
  sh}\xi_{\rm cr}\,\gg\,1$. In the opposite (weak current) limit,
$\gamma_{\rm sh}\xi_{\rm cr}\,\ll\,1$, one finds $\vert\beta_{y,{\rm
    in}\vert{\cal R}}\vert\,\sim\,\gamma_{\rm sh}\xi_{\rm cr}$,
$\beta_{{\cal R}\vert\rm u}\,\sim\,0$ and $\gamma_{{\cal R}\vert\rm
  u}\,\sim\,1$; similarly, $\gamma_{{\cal R}\vert\rm
  sh}\,\sim\,\gamma_{\rm sh}$.

\subsection{Warm plasma limit}\label{sec:app-warm}
The above discussion assumed a cold plasma of returning particles,
with initial momentum (on the shock surface) directed along the shock
normal. Here we introduce the effects of angular dispersion of the
beam of returning particles. For simplicity, we neglect the dispersion
in Lorentz factor of the returning particles; this dispersion can be
taken into account but it should not modify strongly the overall shape
of the current profile.

The number density of returning particles at the shock front
(considering $e^+/e^-$ species altogether), with momentum oriented
within a solid angle element $\boldsymbol{{\rm d}\Omega_{\rm i}}$, is
  written ${\rm d}n_{\rm r+,i}\left(\boldsymbol{\Omega_{\rm
      i}}\right)$. The magnitude of the current deposited by those
  particles in the precursor can be written:
\begin{equation}
{\rm d}j_{y,\rm r+}(x)\,=\,\left\vert\beta_y(x)\right\vert \,{\rm
  d}n_{\rm r+}(x,\boldsymbol{\Omega_{\rm i}})ec\ .\label{eq:jy}
\end{equation}
Assuming that the particle population ${\rm r-}$ deposit the same
amount of current as ${\rm r+}$, and using the equation of
conservation for the number density of ${\rm r+}$ particles, the total
current element deposited by supra-thermal particles emitted in the
$\boldsymbol{\Omega_{\rm i}}$ direction reads:
\begin{equation}
{\rm d}j_{y}(x)\,\simeq\, 2\,\frac{\left\vert\beta_{y,\rm
    r+}(x)\right\vert}{\beta_{x,\rm r+}(x)}\beta_{x,\rm r+,i}{\rm
  d}n_{\rm r+,i}(\boldsymbol{\Omega_{\rm i}})ec\ ,\label{eq:jy}
\end{equation}
$\beta_{x,\rm r+,i}$ denoting the initial $x-$component of the
3-velocity of $\rm r+$ particles.  This equation can be simplified
using the result of the previous section, which indicate that
$\beta_{x,\rm r+}(x)\,\simeq\,\left\vert 2\dot\beta_{x,\rm
  r+}(x_1)\,(x-x_1)\right\vert^{1/2}$ in the vicinity of the turning
point $x_1$, so that most of the current ${\rm d}j_y(x)$ is deposited
at $x_1$. Note that $x_1$ depends on the initial direction
$\boldsymbol{\Omega_{\rm i}}$. We then approximate the spatial profile
of the current element Eq.~(\ref{eq:jy}) with a delta function in $x$:
\begin{equation}
{\rm d}j_y(x)\,\simeq\, {\cal A}\, \,\delta\left(x-x_1\right)\,{\rm
  d}F_{\rm r+,i}(\boldsymbol{\Omega_{\rm i}})\,e \,\label{eq:jy2}
\end{equation}
with ${\rm d}F_{\rm r+,i}(\boldsymbol{\Omega_{\rm i}})\,=\,
\beta_{x,\rm r+,i}c\, {\rm d}n_{\rm r+,i}(\boldsymbol{\Omega_{\rm
    i}})$ the inital flux element. The prefactor is calculated by
normalizing the integrated current element along $x$ in
Eq.~(\ref{eq:jy2}) to that obtained in Eq.~(\ref{eq:jy}). 

In order to express ${\cal A}$ as a function of the initial velocities
$\beta_{x,\rm r+,i}$ and $\beta_{y,\rm r+,i}$, one needs to express
the quantity $\vert\beta_{y,\rm r+}(x)/\beta_{x,\rm r+}(x)\vert$ in
the vicinity of $x_1$ using the equations of motion. These equations
of motion must be written in the upstream rest frame then Lorentz
transformed to the shock frame. We compute the trajectories of the
returning particles in the background electromagnetic field,
neglecting in particular the perturbed component of the magnetic
field; this should remain a good approximation, given that the overall
effect of the angular dispersion of the beam is to spread out over the
precursor length scale the current profile. One then obtains first the
turning point:
\begin{eqnarray}
x_1(\boldsymbol{\Omega_{\rm i}})&\,=\,&\frac{c}{\omega_{\rm c,0}}\,\gamma_{\rm sh}^3\,
\left(1-\beta_{\rm sh}\beta_{x,\rm r+,i}\right)\nonumber\\
&& \times\left[\beta_{x,\rm i\vert u}\sin\varpi_1  +
\beta_{y,\rm i\vert u}\left(1-\cos\varpi_1\right)+\beta_{\rm sh}\varpi_1\right]\ ,\nonumber\\
\end{eqnarray}
as a function of the upstream-frame initial velocities
\begin{equation}
\beta_{x,\rm i\vert u}\,=\,\frac{\beta_{x,\rm r+,i}-\beta_{\rm
    sh}}{1-\beta_{x,\rm r+,i}\beta_{\rm sh}},\quad
\beta_{y,\rm i\vert u}\,=\,\frac{\beta_{y,\rm r+,i}}{\gamma_{\rm
    sh}(1-\beta_{x,\rm r+,i}\beta_{\rm sh})}\ ,
\end{equation}
and the quantity $\varpi_1$, which is defined implicitly by:
\begin{equation}
\beta_{x,\rm i \vert u}\cos\varpi_1 + \beta_{y,\rm i\vert u}\sin\varpi_1\,=\,-\beta_{\rm
  sh}\ .
\end{equation}
Recall that $\beta_{\rm sh}\,<\,0$ in our present notations.  The
initial cyclotron frequency of the returning particles reads
$\omega_{\rm c,0}\,=\,e\gamma_{\rm sh}B_{\rm u}/\left(\gamma_{0,\rm
  r+}m c\right)$, with $\gamma_{0,\rm r+}\simeq \gamma_{\rm sh}$ their
initial Lorentz factor.  One derives eventually:
\begin{eqnarray}
{\cal A}&\,=\,& 2 \sqrt{2}\left(x_1 c/\omega_{\rm
  c,0}\right)^{1/2} \left[\gamma_{\rm sh}(1-\beta_{\rm sh}\beta_{x,\rm
    r+,i})\right]^{1/2}\nonumber\\
&&\quad\times\left\vert(\beta_{x,\rm r+,i}\sin\varpi_1 - \beta_{y,\rm
  r+,i}\cos\varpi_1)\right\vert^{1/2} \ .
\end{eqnarray}
Finally, the flux is normalized through $\int {\rm d}F_{\rm
  r+,i}(\boldsymbol{\Omega_{\rm i}})\,=\,\gamma_{\rm sh}\xi_{\rm
  cr}n_\infty c$.

\begin{figure}
\includegraphics[bb=30 10 330 220, width = 0.49\textwidth]{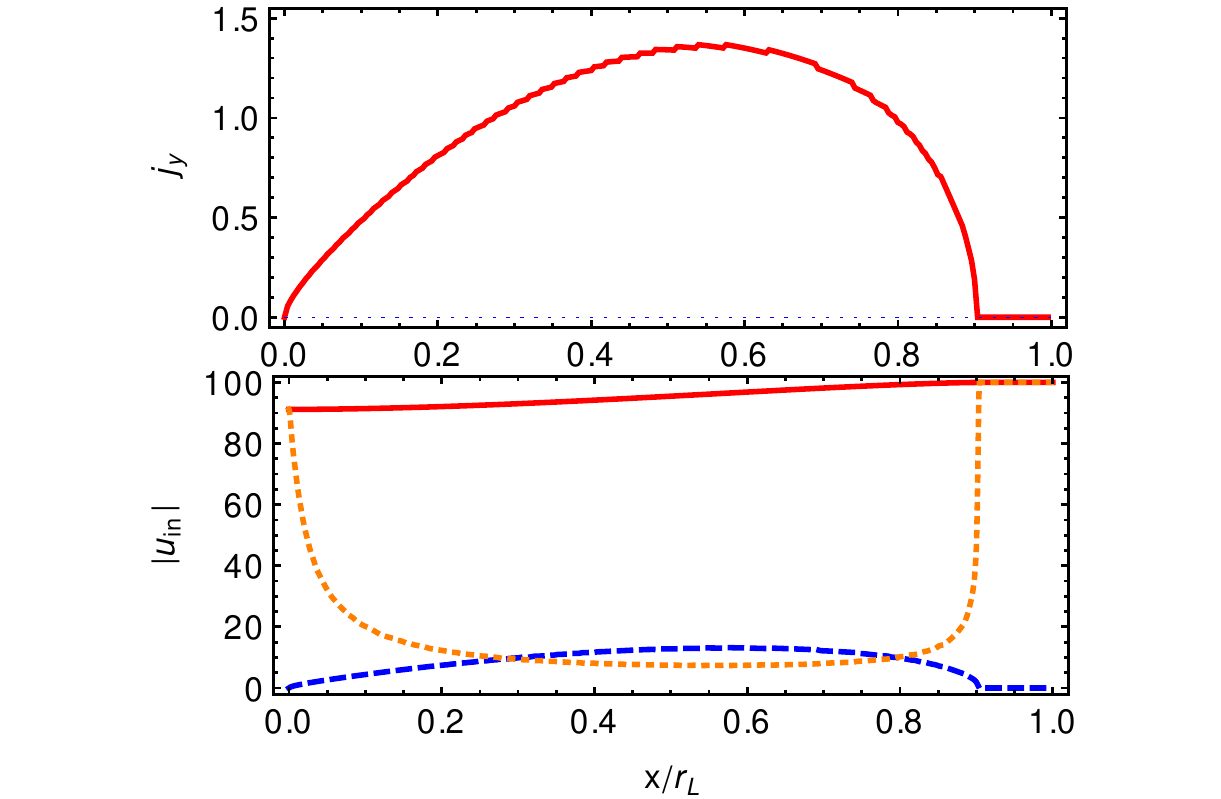}
\caption{Upper panel: profile of $j_y$ (in units of $\gamma_{\rm
    sh}\xi_{\rm cr}n_\infty ec$) carried by returning particles as a
  function of $x/r_{\rm L}$ ($r_{\rm L}=c/\omega_{\rm c}$) in the
  limit $\gamma_{\rm sh}\,\gg\,1$, including the effects of angular
  dispersion at the shock. Lower panel: modulus of the 4-velocity
  components $\vert u_{x,\rm in}\vert$ (solid red), $\vert u_{y,\rm
    in}\vert$ (dashed blue) and $\gamma_{{\cal R}\vert\rm sh}$ (dotted
  orange); the lower panel assumes $\gamma_{\rm sh}=100$ and $\xi_{\rm
    cr}=0.1$.
  \label{fig:prof-warm} }
\end{figure}

In the limit $\gamma_{\rm sh}\,\rightarrow\,+\infty$, all above
quantities reach finite asymptotes, as it should; we use these
asymptotic values in the numerical calculation of the integral over
the angular variables. One finally obtains the current profile
depicted in Fig.~\ref{fig:prof-warm}.

This profile allows to estimate the velocity profile of the incoming
plasma inside the foot. As in the cold plasma limit, current
compensation imposes the following scalings inside the precursor
\begin{equation}
\vert u_{y,\rm in}\vert\,\sim\, \xi_{\rm cr}\gamma_{\rm sh}\ ,
u_{x,\rm in}\,\sim\, -\left(1-\xi_{\rm cr}\right)\gamma_{\rm sh}\ ,
\end{equation}
so that the relative Lorentz factor between the shock front frame and
the frame ${\cal R}$ in which the incoming is at rest along $+x$ is,
as before, $\gamma_{{\cal R}\vert\rm sh}\,\simeq\, 1/\xi_{\rm cr}$ if
$\gamma_{\rm sh}\xi_{\rm cr}\,\gg\,1$. Figure~\ref{fig:prof-warm}
shows a numerical calculation of the evolution of $u_{x,\rm in}$,
$\vert u_{y,\rm in}\vert$ and $\gamma_{{\cal R}\vert\rm sh}$ inside the precursor
(assuming $|b|\,\ll\,1$) for $\gamma_{\rm sh}\,=\,100$ and $\xi_{\rm
  cr}=0.1$, which confirms the above scalings.

\section{Linear system}\label{sec:sys}
We explicit here the linear system used to compute the dispersion
relation, for reference.  We rescale the time and space derivatives by
$\omega_{\rm c}$ (cyclotron frequency in the upstream rest frame):
$\partial_{\tilde t}\,\equiv\,\omega_{\rm c}^{-1}\partial_t$,
$\partial_{\tilde x}\,\equiv\,c\omega_{\rm c}^{-1}\partial_x$ etc.  We
rescale all electromagnetic fields by the background value $B_{z{\cal
    R}}$ (e.g. $\delta \tilde B_x\,\equiv, \delta B_x/B_{z{\cal R}}$)
and we introduce the notations: $\kappa\,\equiv\,\gamma_{{\cal
    R}\vert\rm sh}/u^0$, $\tilde\beta_{\rm s}^2 \,\equiv\,\beta_{\rm
  s}^2/u^{0\,2}$, $\delta_n\,\equiv\, \delta n/n$,
$\delta_\rho\,\equiv\,\delta \rho/n$, and we rescale $\delta u^\mu$
and $\Delta u^\mu$ by $u^0$, e.g.  $\delta \tilde u^\mu\,\equiv\,
\delta u^\mu/u^0$.  This leads to the following adimensioned system
\begin{eqnarray}
  \partial_{\tilde t} \delta_n + \beta_y\partial_{\tilde y}\delta_\rho +  
  \beta_y\partial_{\tilde t} \Delta \tilde u_y +\partial_{\tilde x} \delta \tilde u_x
  + \partial_{\tilde y} \delta \tilde u_y
  + \partial_{\tilde z}\delta \tilde u_z &\,=\,&0\nonumber\\
  \partial_{\tilde t} \delta_\rho + \beta_y\partial_{\tilde y}\delta_n + 
  \beta_y\partial_{\tilde t}\delta \tilde u_y + \partial_{\tilde x}\Delta \tilde u_x
  +\partial_{\tilde y}\Delta \tilde u_y + \partial_{\tilde z}\Delta \tilde u_z &\,=\,&0\nonumber\\
  \partial_{\tilde t} \delta \tilde u_x +\beta_y\partial_{\tilde y} \Delta \tilde u_x +
  \tilde\beta_{\rm s}^2 \partial_{\tilde x}\delta_n - \kappa \Delta \tilde u_y -
  \beta_y\kappa \delta \tilde B_z&\,=\,&0\nonumber\\
  \partial_{\tilde t} \Delta \tilde u_x + \beta_y\partial_{\tilde y} \delta \tilde u_x +
  \tilde\beta_{\rm s}^2 \partial_{\tilde x}\delta_\rho - \kappa \delta \tilde u_y -
  \kappa \delta \tilde E_x&\,=\,&0\nonumber\\
  \partial_{\tilde t}\delta \tilde u_y  + \beta_y\partial_{\tilde y} \Delta \tilde u_y +
  \tilde\beta_{\rm s}^2 \partial_{\tilde y}\delta_n + \beta_{{\cal R}\vert\rm u}\kappa\beta_y \delta \tilde u_y + \kappa \Delta \tilde u_x  &\,=\,&0\nonumber\\
  \partial_{\tilde t}\Delta \tilde u_y + \beta_y\partial_{\tilde y} \delta \tilde u_y +
  \tilde\beta_{\rm s}^2 \partial_{\tilde y}\delta_\rho +\beta_{{\cal R}\vert\rm
    u}\kappa\beta_y \Delta \tilde u_y && \nonumber\\ - \kappa \delta \tilde E_y + \kappa \delta \tilde u_x  &\,=\,&0\nonumber\\
  \partial_{\tilde t} \delta \tilde u_z + \beta_y\partial_{\tilde y} \Delta \tilde u_z + \tilde\beta_{\rm s}^2 \partial_{\tilde z}\delta_n +
  \beta_y\kappa \delta \tilde B_x&\,=\,&0\nonumber\\
  \partial_{\tilde t} \Delta \tilde u_z + \beta_y\partial_{\tilde y} \delta \tilde u_z + 
  \tilde\beta_{\rm s}^2 \partial_{\tilde z}\delta_\rho -
  \kappa \delta \tilde E_z&\,=\,&0\nonumber\\
  \partial_{\tilde t} \delta \tilde B_x + \partial_{\tilde y} \delta \tilde E_x - \partial_{\tilde z}
  \delta \tilde E_y&\,=\,&0\nonumber\\
  \partial_{\tilde t} \delta \tilde B_y + \partial_{\tilde z}\delta \tilde E_x - \partial_{\tilde x}\delta
  \tilde E_z&\,=\,&0\nonumber\\
  \partial_{\tilde t} \delta \tilde B_z + \partial_{\tilde x}\delta \tilde E_y - \partial_{\tilde y} \delta
  \tilde E_x&\,=\,&0\nonumber\\
  \partial_{\tilde t} \delta \tilde E_x - \partial_{\tilde y}\delta \tilde B_z + \partial_{\tilde z}\delta
  \tilde B_y +\frac{1}{\kappa \sigma}\Delta \tilde u_x&\,=\,&0 \nonumber\\
  \partial_{\tilde t} \delta \tilde E_y - \partial_{\tilde z}\delta \tilde B_x
  + \partial_{\tilde x}\delta \tilde B_z + \frac{1}{\sigma}\Delta \tilde u_y +
  \frac{1}{\kappa \sigma}\beta_y\delta_n&\,=\,&0\nonumber\\
  \partial_{\tilde t}\delta \tilde E_z - \partial_{\tilde x}\delta \tilde B_y + \partial_{\tilde y}\delta
  \tilde B_x + \frac{1}{\kappa \sigma}\Delta \tilde u_z&\,=\,&0\nonumber\\
  &&
\end{eqnarray}

\end{document}